\documentclass[12pt]{article}
\usepackage{fullpage}
\usepackage{cite}
\usepackage{amsmath}
\usepackage{amssymb}
\usepackage{graphicx}
\usepackage[utf8]{inputenc} 

\title{}
\date{}
\def\para{\\ [-2mm]}

\def\para{\\ [-2mm]}
\def \be  {\begin{equation}}
\def \ee  {\end{equation}}
\def \ba  {\begin{eqnarray}}
\def \ea  {\end{eqnarray}}
\newcommand{\nn}{\nonumber}
\def\eqn#1{eq.~(\ref{#1})} 
\def\eqns#1#2{eqs.~(\ref{#1}) and~(\ref{#2})}

\def\IZ{\relax\ifmmode\mathchoice
{\hbox{\cmss Z\kern-.4em Z}}{\hbox{\cmss Z\kern-.4em Z}}
{\lower.4pt\hbox{\cmsss Z\kern-.4em Z}}
{\lower1.2pt\hbox{\cmsss Z\kern-.4em Z}}\else{\cmss Z\kern-.4em Z}\fi}
\newcommand{\Z}{\mathsf{Z}\kern -5pt \mathsf{Z}}
\newcommand{\unit}{\mathsf{1}\kern -3pt \mathsf{l}}

\def \Tr {\mathop{\rm Tr}\nolimits}

\def\half{{\textstyle{1 \over 2}}}
\def\fr#1#2{ {\textstyle{#1 \over #2}}}

\def\de {\epsilon}

\def\eps{\epsilon}

\def\cA {  {\cal A} }

\def\cI {  {\cal I} }

\def\cM {  {\cal M} }
\def\cN {  {\cal N} }
\def\cO {  {\cal O} }
\def\< { \langle}
\def\> { \rangle}

\def\atil{\tilde{a}}
\def\etatil{\tilde{\eta}}

\def\bone{1\kern -3pt \mathrm{l}}

\def\cG {  {\cal G} }
\def\cH {  {\cal H} }
\def\BX { B }
\def\hatBX { \hat \BX }
\def\hatcA { \hat \cA }
\def\hatA { \hat A }
\def\CX { C }
\def\hX { h }
\def\ik  { {i k} }
\def\ellk  { {\ell k} }
\def\one   { {00} }
\def\two   { {11} }
\def\three { {21} }
\def\four  { {22} }
\def\five  { {31} }
\def\six   { {32} }
\def\seven { {33} }
\def\BXL { \BX_{\ell\ell}^\Ell }

\def\Zero{ { (0) }}
\def\One {{ (1) }}
\def\Two{{(2)}}
\def\Three{{(3)}}

\def\lam{\lambda}

\def\e { {\rm e} }

\def\t { t }
\def\tf{ \tilde{f} }
\def\bT{\mathbf{T}}

\def\bS{\mathbf{S}}
\def\bW{\mathbf{W}}
\def\bGam{\mathbf{\Gamma}}
\def\suml{\sum_{\ell=1}^\infty}
\def\Ell{{(\ell)}}
\def\Ellk{{(\ell,k)}}
\def\Elltwok{{(\ell,2k)}}
\def\Elltwokplus{{(\ell,2k+1)}}

\begin{document}

\titlepage
\begin{flushright}
BOW-PH-169\\
\end{flushright}

\vspace{3mm}

\begin{center}

{\Large\bf\sf
All-loop-orders relation between Regge limits of 
\\ [2mm]
$\cN=4$ SYM 
and $\cN=8$ supergravity 
\\ [4mm]
four-point amplitudes
}

\vskip 3cm

{\sc
Stephen G. Naculich
}

\vskip 0.5cm
{\it
Department of Physics\\
Bowdoin College\\
Brunswick, ME 04011 USA
}

\vspace{5mm}
{\tt
naculich@bowdoin.edu
}
\end{center}

\vskip 3cm

\begin{abstract}

We examine in detail the structure of the Regge limit 
of the (nonplanar) $\cN=4$ SYM four-point amplitude.
We begin by developing a basis of color factors $\CX_\ik$ 
suitable for the Regge limit of the amplitude at any loop order, 
and then calculate explicitly the coefficients of the amplitude 
in that basis through three-loop order using the Regge limit 
of the full amplitude previously calculated by Henn and Mistlberger.
We compute these coefficients exactly at one loop, 
through $\cO(\de^2)$ at two loops, 
and through $\cO(\de^0)$ at three loops,
verifying that the IR-divergent pieces are consistent
with (the Regge limit of) the expected infrared divergence structure, 
including a contribution from the three-loop correction 
to the dipole formula.
We also verify consistency with the IR-finite NLL and NNLL 
predictions of Caron-Huot et al. 
Finally we use these results to motivate the conjecture 
of an all-orders relation between one of the coefficients 
and the Regge limit of the $\cN=8$ supergravity four-point amplitude. 
\para

\end{abstract}

\vspace*{0.5cm}

\vfil\break

\section{Introduction} 
\setcounter{equation}{0}
\label{sec:intro}

Maximally supersymmetric Yang-Mills and supergravity theories
have garnered especial interest 
ever since it was shown that 
they emerge from the low energy limit of superstring theory,
which allowed the first one-loop calculation
of four-point amplitudes in each \cite{Green:1982sw}.
In the decades following,
higher-loop $\cN=4$ SYM four-point amplitudes were 
calculated \cite{Bern:1997nh,Bern:1998ug,Bern:2007hh,Bern:2008pv,Bern:2010tq,
Bern:2012uf,Bern:2012uc}
in terms of planar and nonplanar scalar integrals 
through the use of generalized unitarity \cite{Bern:1994zx,Bern:1994cg}.
The infrared divergences of these massless integrals
were dimensionally regulated in $D=4-2\de$ dimensions,
and Laurent expansions in $\de$ of the planar 
two- and three-loop integrals \cite{Smirnov:1999gc,Smirnov:2003vi}
were used to obtain explicit expressions 
for planar (leading in the $1/N$ expansion) 
$\cN=4$ SYM four-point amplitudes \cite{Anastasiou:2003kj,Bern:2005iz}.
Using these results,
together with the well-studied structure of infrared divergences of gauge 
theories \cite{Catani:1998bh,Sterman:2002qn},
Bern, Dixon, and Smirnov proposed their celebrated all-loop-orders
ansatz for (maximally-helicity-violating) planar $n$-point
amplitudes \cite{Bern:2005iz},
whose validity for $n=4$ and 5 follows from dual conformal
invariance of the planar theory \cite{Drummond:2007cf,Drummond:2007au}.
Progress on the evaluation of nonplanar two- and three-loop integrals
has taken longer \cite{Tausk:1999vh,Henn:2013tua,Henn:2013nsa,Henn:2020lye},
but these results have allowed explicit expressions 
for the full (nonplanar) $\cN=4$ SYM four-point amplitude
at two \cite{Naculich:2008ys,Naculich:2013xa}
and, more recently, three \cite{Henn:2016jdu} loops.
These expressions were verified to be consistent 
with the expected (nonplanar) infrared divergence structure 
through three loops \cite{Aybat:2006wq,Aybat:2006mz,
Dixon:2008gr,Becher:2009cu,Gardi:2009qi,Dixon:2009gx,
Becher:2009qa,Gardi:2009zv,Dixon:2009ur,Almelid:2015jia}.
\para

Higher-loop four-point amplitudes of $\cN=8$ supergravity, 
a theory believed to be ultraviolet finite to a high loop order
(see ref.~\cite{Bern:2018jmv} and references therein),
have also been calculated in terms of planar and nonplanar integrals
\cite{Bern:1998ug,Bern:2007hh,Bern:2012uf,Bern:2017ucb}.
Again using Laurent expansions of the planar and nonplanar integrals,
explicit expressions for the integrated 
$\cN=8$ supergravity four-point amplitudes 
have been obtained at 
two \cite{Naculich:2008ew,Brandhuber:2008tf,BoucherVeronneau:2011qv}
and, more recently, three \cite{Henn:2019rgj} loops.
\para

Relations between supergravity and subleading-color SYM amplitudes 
and their IR divergences were explored in 
refs.~\cite{Naculich:2008ys,Bern:2011rj,Naculich:2011fw,
BoucherVeronneau:2011qv,Naculich:2011my,Naculich:2013xa}.
Exact relations between $\cN=4$ SYM and $\cN=8$ supergravity 
four-point amplitudes were established 
at one- and two-loop levels \cite{Naculich:2008ys},
but attempts at finding relations at three loops and beyond were
generally unsuccessful.
\para

Scattering amplitudes often exhibit dramatic simplification
in the high-energy (or Regge) limit $s \gg -t$.
The known structure of infrared divergences of gauge theory amplitudes
\cite{Catani:1998bh,Sterman:2002qn,Bern:2005iz,
Aybat:2006wq,Aybat:2006mz,
Dixon:2008gr,Becher:2009cu,Gardi:2009qi,Dixon:2009gx,
Becher:2009qa,Gardi:2009zv,Dixon:2009ur,Almelid:2015jia}
simplifies at high energies, giving rise to 
leading and subleading logarithmic behavior
\cite{Bret:2011xm,DelDuca:2011ae,DelDuca:2013ara,DelDuca:2014cya}.
Moreover, 
an effective Hamiltonian approach 
based on Balitsky-Fadin-Kuraev-Lipatov theory 
was used to compute IR-finite large logarithmic behavior
(NLL and NNLL) in the Regge limit 
\cite{Caron-Huot:2013fea,Caron-Huot:2017fxr,Caron-Huot:2017zfo,Caron-Huot:2020grv}.
The Regge limit of the $\cN=4$ SYM four-point amplitude was examined in 
refs.~\cite{Drummond:2007aua,Naculich:2007ub,DelDuca:2008pj,
Naculich:2009cv,Henn:2010bk,Henn:2010ir,Henn:2016jdu,Caron-Huot:2017fxr}.
\para

The high-energy limit of gravity amplitudes 
is in some ways even simpler than gauge theory amplitudes
\cite{tHooft:1987vrq,Muzinich:1987in,Amati:1987wq,
Amati:1987uf,Amati:1990xe,Amati:1992zb,Amati:1993tb,
Verlinde:1991iu,Kabat:1992tb}.
The high-energy behavior of $\cN=8$ supergravity amplitudes 
has been studied in refs.~\cite{Grisaru:1981ra,Grisaru:1982bi,
Schnitzer:2007kh,Schnitzer:2007rn,
Giddings:2010pp,
BoucherVeronneau:2011qv,
Bartels:2012ra,
Melville:2013qca,Henn:2019rgj},
and recently, 
eikonal exponentiation in impact parameter space 
has been used to obtain an exact all-loop-orders expression 
for the high-energy limit 
of the $\cN=8$ supergravity four-point amplitude
\cite{DiVecchia:2019myk,DiVecchia:2019kta}.
\para

In this paper, we examine in detail the structure 
of the Regge limit of the (nonplanar) $\cN=4$ SYM four-point amplitude.
We begin by developing a basis of color factors $\CX_\ik$ 
suitable for the Regge limit of the amplitude 
at any loop order, 
and then calculate explicitly the coefficients of the 
amplitude in that basis through three-loop order
using the Regge limit of the full amplitude previously calculated 
by Henn and Mistlberger \cite{Henn:2016jdu}.
We compute these coefficients exactly at one loop, 
through $\cO(\de^2)$ at two loops, 
and through $\cO(\de^0)$ at three loops,
verifying that the IR-divergent pieces are consistent
with (the Regge limit of) the expected infrared divergence 
structure as determined in 
refs.~\cite{Catani:1998bh,Sterman:2002qn,Bern:2005iz,
Aybat:2006wq,Aybat:2006mz,
Dixon:2008gr,Becher:2009cu,Gardi:2009qi,Dixon:2009gx,
Becher:2009qa,Gardi:2009zv,Dixon:2009ur,Almelid:2015jia,
Bret:2011xm,DelDuca:2011ae,Caron-Huot:2013fea,
DelDuca:2013ara,DelDuca:2014cya,Caron-Huot:2017fxr},
including a contribution from the three-loop correction 
to the dipole formula \cite{Almelid:2015jia,Caron-Huot:2017fxr}.
We also verify consistency\footnote{The authors of 
ref.~\cite{Caron-Huot:2017fxr} also reported consistency of 
their results with ref.~\cite{Henn:2016jdu}.}
with the IR-finite NLL and NNLL predictions
of Caron-Huot et al. \cite{Caron-Huot:2013fea,Caron-Huot:2017fxr}.
Finally we use these results to motivate the 
conjecture of an all-orders relation 
between one of the coefficients and the
Regge limit of the $\cN=8$ supergravity four-point amplitude. 
\para

Color-ordered $n$-gluon amplitudes 
(i.e. the coefficients of the amplitude in a trace basis)
of an SU($N$) gauge theory
are not independent but obey relations derived from group 
theory \cite{Green:1982sw,Mangano:1990by,Bern:1990ux,Bern:2002tk,Naculich:2011ep,Edison:2011ta,Edison:2012fn}.
For four-gluon amplitudes, 
there are $3\ell -1$ independent 
color-ordered amplitudes at $\ell \ge 2$ loops
(with two at tree level, and three at one loop).
In the Regge limit, each of these is reduced by one.
Inspired by refs.~\cite{DelDuca:2011ae,Caron-Huot:2013fea,DelDuca:2014cya,Caron-Huot:2017fxr},
we define a set of color factors $\CX_\ik$ by
\begin{align}
\CX_\one 
&= \tf^{a_4 a_1 e} \tf^{a_2 a_3 e}  \,,
&&\nn\\[2mm]
\CX_{ii} 
&=   (\bT_{s-u})^i \CX_\one \,, 
&& i \ge 1 
\nn\\[2mm]
\CX_{i 1} 
&=   [\bT_t^2, \cdots, [\bT_t^2,  [\bT_t^2,  \bT_{s-u}^2]]\cdots ] \CX_\one \,, && i \ge 2
\nn\\[2mm]
\CX_{i, i-1} 
&=   [\bT_{s-u}^2, \cdots, [\bT_{s-u}^2, [\bT_t^2,  \bT_{s-u}^2]]\cdots ] \CX_\one \,, 
&& i \ge 2
\label{defCik}
\end{align}
where $\bT^2_t$ and $\bT^2_{s-u}$ are operators (cf. \eqn{keyobs}) 
that act on the tree-level $t$-channel color factor $\CX_\one$
to generate higher-loop color factors.
Each color factor $\CX_\ik$ in \eqn{defCik} 
contains exactly $i$ operators $\bT^2_t$ or $\bT^2_{s-u}$
with $k$ factors of $\bT_{s-u}^2$.
These color factors have well-defined signature (either $+$ or $-$)
under the crossing symmetry operation 
that exchanges external legs
2 and 3 (i.e., $s \leftrightarrow u$).
As we will see in sec.~\ref{sec:IR},
this basis is particularly well-suited 
to describe the IR divergences of the amplitude in the Regge limit.
\para

Having defined the Regge basis of color factors, we write 
the $\cN=4$ SYM four-point amplitude in the Regge limit as
\be
\cA
= A_1^\Zero 
\sum_{\ell=0}^\infty
\atil^\ell
\sum_{i=0}^\ell \sum_{k}
\BX_\ik^\Ell N^{\ell-i}   \CX_\ik
\label{reggedecomp}
\ee

where $\atil$ is the loop expansion parameter,
and the range of $k$ is
\be
k = 
\begin{cases}
0, & \mbox{when } i=0, \\
1, & \mbox{when } i=1,  \\
1, 2, & \mbox{when } i=2, \\
1, i-1, i,  & \mbox{when } i \ge 3. 
\end{cases}
\ee

By using the decomposition of the color factors $\CX_\ik$
into a trace basis, the coefficients $\BX_\ik^\Ell$ 
can be expressed as linear combinations of color-ordered amplitudes.
By Bose symmetry of the full amplitude, 
the coefficients $\BX^\Ell_\ik$ will have
the same signature under crossing symmetry as their associated
color factors $\CX_\ik$,
which has implications \cite{Caron-Huot:2017fxr} 
for the reality properties of $\BX_\ik^\Ell$,
as we will see in this paper.
\para

By examining the Regge limit of the structure of IR divergences 
for $\cN=4$ SYM theory, 
we determine that the $\BX^\Ell_\ik$ are polynomials 
of order $\ell-k$ in $ L \equiv \log |s/t| - \half i \pi$.
Thus the leading logarithmic (LL) behavior of the amplitude at $\ell$ loops
is entirely contained in the coefficient $\BX_\one^\Ell$,
with the other coefficients contributing to 
subleading logarithmic behavior.
\para

The full amplitude $\cA$ can be factored
into an infrared-divergent prefactor 
times an infrared-finite hard function $\cH$,
whose coefficients in the Regge basis of color factors
are denoted $\hX_\ik^\Ell$, analogous to \eqn{reggedecomp}. 
Using the Regge limit of the $\cN=4$ SYM color-ordered amplitudes
given in ref.\cite{Henn:2016jdu},
we compute the IR-finite coefficients $\hX_\ik^\Ell$
exactly at one loop, 
through $\cO(\de^2)$ at two loops, 
and through $\cO(\de^0)$ at three loops.
The NLL and NNLL contributions to $\hX_\ik^\Ell$
are verified to agree with the predictions of 
refs.~\cite{Caron-Huot:2013fea,Caron-Huot:2017fxr}.
\para

Finally, using our results through three loops, 
we observe that one of the coefficients, $\BXL$, 
which contains no $\log|s/t|$ dependence, 
is  equal (up to a sign) to the Regge limit of the 
$\ell$-loop $\cN=8$ supergravity four-point amplitude.
At one and two loops, this equality follows from the
exact relations presented in refs.~\cite{Naculich:2008ys}.
Moreover, this relation is consistent with the leading
IR divergences of the amplitudes for all $\ell$.  
We therefore conjecture that this SYM-supergravity relation 
holds to all loop orders in the Regge limit.
\para

This paper is structured as follows.
We review the group theory constraints on four-gluon amplitudes
in sec.~\ref{sec:constraints},
and the Regge limits of the tree-level and one-loop
$\cN=4$ SYM amplitudes in sec.~\ref{sec:oneandtwoloop}.
In sec.~\ref{sec:reggebasis}, we develop a basis of color factors $\CX_\ik$
suitable for the Regge limit of the $\cN=4$ SYM amplitude at any loop order.
We then obtain expressions (through three loops)
for the coefficients $\BX_\ik^\Ell$ in this basis 
as linear combinations of color-ordered amplitudes
(and for all loops for $\BXL$).
In sec.~\ref{sec:IR},
after reviewing the structure of (the Regge limit of) 
the infrared divergences 
of the $\cN=4$ SYM amplitude through three loops,
we compute the infrared-finite hard function
using the results of ref.~\cite{Henn:2016jdu}.
In sec.~\ref{sec:grav},
we review the Regge limit of the 
$\cN=8$ supergravity four-graviton amplitude,
and conjecture its equivalence to $\BXL$.
In sec.~\ref{sec:concl}, we offer some concluding remarks.

\section{Group-theory constraints on four-point amplitudes} 
\setcounter{equation}{0}
\label{sec:constraints}

In this section, we briefly review the group-theory constraints 
on the $\ell$-loop color-ordered four-gluon amplitudes
in an SU($N$) gauge theory that were derived in ref.~\cite{Naculich:2011ep}.
These will play an important role in the following sections
when we define a new basis of color factors
for four-gluon amplitudes in the Regge limit.
\para

The color-ordered amplitudes of a gauge theory are
the coefficients of the full amplitude in a basis
using traces of generators $T^a$ in the fundamental representation
of the gauge group\footnote{Our conventions are 
$\Tr(T^a T^b) = \delta^{ab}$ so that 
$[T^a, T^b] = i \sqrt2 f^{abc} T^c$ and 
$f^{abc} = (-i/\sqrt2) \Tr([T^a, T^b] T^c)$.
\label{Tconvention} }.
Color-ordered amplitudes
have the advantage of being individually gauge-invariant.
Four-gluon amplitudes of SU($N$) gauge theories can be
expressed in terms of a six-dimensional basis 
$c_{[\lam]}$ of single and double 
traces
\cite{Bern:1990ux}
\begin{align}
c_{[1]} &=  
\Tr(T^{a_1} T^{a_2} T^{a_3} T^{a_4}) + \Tr(T^{a_1} T^{a_4} T^{a_3} T^{a_2}),
\qquad
c_{[4]} =  \Tr(T^{a_1} T^{a_3}) \Tr(T^{a_2} T^{a_4}) , 
\nn\\
c_{[2]} &= 
\Tr(T^{a_1} T^{a_2} T^{a_4} T^{a_3}) + \Tr(T^{a_1} T^{a_3} T^{a_4} T^{a_2}),
\qquad
c_{[5]}=  \Tr(T^{a_1} T^{a_4}) \Tr(T^{a_2} T^{a_3})  ,
\label{sixdimbasis} 
\\
c_{[3]} &=  
\Tr(T^{a_1} T^{a_4} T^{a_2} T^{a_3}) + \Tr(T^{a_1} T^{a_3} T^{a_2} T^{a_4}),
\qquad
c_{[6]} =  \Tr(T^{a_1} T^{a_2}) \Tr(T^{a_3} T^{a_4}) . \nn
\end{align}

All other possible trace terms vanish in SU($N$) since $\Tr(T^a)=0$.
The $\ell$-loop color-ordered amplitudes can be
further decomposed \cite{Bern:1997nh}
in powers of $N$ as
\be
\cA^\Ell =
\sum_{\lam = 1}^3
\left( \sum_{k=0}^{\lfloor \frac{\ell}{2}  \rfloor} 
N^{\ell-2k} A^\Elltwok_\lam \right) c_{[\lam]}
+ \sum_{\lam = 4}^6
\left( \sum_{k=0}^{\lfloor \frac{\ell-1}{2}  \rfloor}
N^{\ell-2k-1} A^\Elltwokplus_\lam\right) c_{[\lam]}
\label{decomp}
\ee

where  $A^{(\ell,0)}_\lam$ are leading-order-in-$N$ (planar) amplitudes,
and  $A^\Ellk_\lam$, $k = 1, \cdots, \ell$,  are subleading-order,
yielding $(3\ell+3)$ color-ordered amplitudes at $\ell$ loops.
The $1/N$ expansion in \eqn{decomp}
suggests an enlargement of the trace basis to
an extended $(3\ell+3)$-dimensional trace basis $ t^\Ell_\lam $,
defined by 
\begin{align}
t^\Ell_{1+ 6k} &=   N^{\ell - 2k} \, c_{[1]} \,,
\qquad\qquad\qquad t^\Ell_{4+6k} =  N^{\ell - 2k - 1} \, c_{[4]} \,,
\nn \\
t^\Ell_{2+ 6k} &=  N^{\ell-2k} \, c_{[2]} \,,
\qquad\qquad\qquad t^\Ell_{5+6k}  = N^{\ell - 2k - 1}  \, c_{[5]} \,,
\label{extendedbasis}
\\
t^\Ell_{3+ 6k} &=  N^{\ell-2k} \, c_{[3]} \,,
\qquad\qquad\qquad t^\Ell_{6 +6k}=  N^{\ell - 2k - 1} \, c_{[6]} \,,
\nn
\end{align} 

in terms of which \eqn{decomp} becomes
\be
\cA^\Ell = \sum_{\lam =1}^{3\ell+3}  A^\Ell_\lam t^\Ell_\lam,
\qquad
{\rm where}
\qquad
A^\Ell_{\lam + 6k}=
\begin{cases}
A^\Elltwok_\lam ,    & \lam = 1, 2, 3 \,,\\
A^\Elltwokplus_\lam,  & \lam = 4, 5, 6 \,.
\end{cases}
\label{expandeddecomp}
\ee

The $(3\ell+3)$ components $A^\Ell_\lam$ are not independent
but are related by various group-theory constraints.
These constraints can be conveniently expressed \cite{Naculich:2011ep}
in terms of a set of null vectors in the space spanned by $t^\Ell_\lam$.
Each $\ell$-loop null vector $r^\Ell_\lam$
gives rise to a linear constraint on the components 
$A^\Ell_\lam$, namely
\be
0= 
\sum_{\lam =1}^{3\ell+3}  
r^\Ell_\lam A^\Ell_\lam \,.
\ee

At tree level, the single null vector
\be
r^\Zero = (1, 1, 1)
\label{treelevelnull}
\ee
corresponds to the constraint \cite{Green:1982sw,Mangano:1990by}
\be
0 = A_1^\Zero + A_2^\Zero + A_3^\Zero \,.
\label{treelevelgrouptheory}
\ee

At one loop, there are three null vectors 
\be
r^\One = 
\left\{ 
\begin{array}{rrrrrr}
(6,& 6,& 6,& -1,& -1,& -1), \\
(0,& 0,& 0,&  1,& -2,&  1), \\
(0,& 0,& 0,&  1,&  0,& -1). 
\end{array}
\right.
\label{oneloopnull}
\ee

The corresponding constraints
imply that the three one-loop subleading-color amplitudes are equal 
and proportional to the sum of leading-color amplitudes \cite{Bern:1990ux}
\be
A^\One_4 = A^\One_5 = A^\One_6 = 2 (A^\One_1 + A^\One_2 + A^\One_3)  \,.
\label{oneloopgrouptheory}
\ee

There are exactly four null vectors for every loop level two and above
\cite{Naculich:2011ep}. 
The four null vectors at two loops are
\be
r^\Two = 
\left\{ 
\begin{array}{rrrrrrrrrr}
(6,& 6,& 6,& -1,& -1,& -1,& 0,& 0,& 0), \\
(0,& 0,& 0,&  1,& -2,&  1,& 1,&-2,& 1), \\
(0,& 0,& 0,&  1,&  0,& -1,& 1,& 0,&-1), \\
(0,& 0,& 0,&  0,&  0,&  0,& 1,& 1,& 1),
\end{array}
\right.
\label{twoloopnull}
\ee

which give rise to four two-loop group-theory relations \cite{Bern:2002tk}.
Consequently, the two-loop amplitude may be written in 
terms of five independent components
$A_1^\Two, A_2^\Two, A_3^\Two, A_7^\Two$, $ A_9^\Two$,
with the four remaining color-ordered amplitudes given by
\begin{align}
A^\Two_4  &=  2(A^\Two_1 +A^\Two_2 +  A^\Two_3) - A^\Two_7  \,,\nn\\
A^\Two_5  &= 2(A^\Two_1 +A^\Two_2 +  A^\Two_3) +  A^\Two_7 + A^\Two_9 \,,\nn\\
A^\Two_6  &=  2(A^\Two_1 +A^\Two_2 +  A^\Two_3)- A^\Two_9   \,, \nn\\
A^\Two_8  &=  - A^\Two_7 - A^\Two_9\,. 
\label{twoloopgrouptheory}
\end{align}

The four null vectors at three loops are
\be
r^\Three = 
\left\{ 
\begin{array}{rrrrrrrrrrrr}
(6,& 6,& 6,&-1,&-1,&-1,& 2,& 2,& 2,& 0,& 0,& 0), \\
(0,& 0,& 0,& 0,& 0,& 0,& 6,& 6,& 6,&-1,&-1,&-1), \\
(0,& 0,& 0,& 0,& 0,& 0,& 0,& 0,& 0,& 1,&-2,& 1), \\
(0,& 0,& 0,& 0,& 0,& 0,& 0,& 0,& 0,& 1,& 0,&-1). 
\end{array}
\right.
\label{threeloopnull}
\ee
As a result, 
the three-loop amplitude can be written in terms of
eight independent components
$A_1^\Three, A_2^\Three, A_3^\Three, A_4^\Three, A_6^\Three,
A_7^\Three, A_8^\Three,$ $ A_9^\Three$,
with the four remaining color-ordered amplitudes given by
\begin{align}
A^\Three_{5} &= 
 6( A^\Three_1 +A^\Three_2 + A^\Three_3)
- A^\Three_4 - A^\Three_6
+ 2( A^\Three_7 +A^\Three_8 +A^\Three_9)\,, \nn\\
A^\Three_{10} &=  2( A^\Three_7 +A^\Three_8 +A^\Three_9) \,, \nn\\
A^\Three_{11} &=  2( A^\Three_7 +A^\Three_8 +A^\Three_9) \,, \nn\\
A^\Three_{12} &=  2( A^\Three_7 +A^\Three_8 +A^\Three_9)\,. 
\label{threeloopgrouptheory}
\end{align}

For all even loop levels beyond two ($\ell = 2 m + 2$),
the four null vectors are given by 
prepending $6m$ zeros to each of the two-loop null vectors (\ref{twoloopnull}),
which give rise to the constraints
\begin{align}
A_{3 \ell-2}^\Ell
&=  2 (A_{3 \ell-5}^\Ell+A_{3 \ell-4}^\Ell+A_{3 \ell-3}^\Ell)
-A_{3 \ell+1}^\Ell
\,, &&\nn\\
A_{3 \ell-1}^\Ell
&=  2 (A_{3 \ell-5}^\Ell+A_{3 \ell-4}^\Ell+A_{3 \ell-3}^\Ell)
+A_{3 \ell+1}^\Ell+A_{3 \ell+3}^\Ell
\,, &&
\nn\\
A_{3 \ell}^\Ell
&=  2 (A_{3 \ell-5}^\Ell+A_{3 \ell-4}^\Ell+A_{3 \ell-3}^\Ell)
-A_{3 \ell+3}^\Ell
\,, &&
\hbox{for even $\ell \ge 2$}
\nn\\
A_{3 \ell+2}^\Ell
&=  -A_{3 \ell+1}^\Ell-A_{3 \ell+3}^\Ell \,.
&& 
\end{align}

For all odd loop levels beyond three ($\ell = 2m+3$), 
the four null vectors are given by prepending $6m$ zeros 
to each of the three-loop null vectors (\ref{threeloopnull})
giving rise to the constraints
\begin{align}
A_{3 \ell-4}^\Ell
&=  6 (A_{3 \ell-8}^\Ell+A_{3 \ell-7}^\Ell+A_{3 \ell-6}^\Ell)
-A_{3 \ell-5}^\Ell-A_{3 \ell-3}^\Ell
+2 (A_{3 \ell-2}^\Ell+A_{3 \ell-1}^\Ell+A_{3 \ell}^\Ell) \,,
&& \nn\\
A_{3 \ell+1}^\Ell
&=  2 (A_{3 \ell-2}^\Ell+A_{3 \ell-1}^\Ell+A_{3 \ell}^\Ell)
\,, && \nn\\
A_{3 \ell+2}^\Ell
&=  2 (A_{3 \ell-2}^\Ell+A_{3 \ell-1}^\Ell+A_{3 \ell}^\Ell)
\,, 
&& \hbox{for odd $\ell \ge 3$}
\nn\\
A_{3 \ell+3}^\Ell
&=  2 (A_{3 \ell-2}^\Ell+A_{3 \ell-1}^\Ell+A_{3 \ell}^\Ell) \,.
&& 
\end{align}
Using \eqn{expandeddecomp},
these constraints can be rewritten in terms of the 
color-ordered amplitudes $A_\lam^{(\ell,k)}$, 
in which case it becomes evident that they involve
only the two or three most-subleading-color amplitudes
(i.e., subleading in the $1/N$ expansion).

\section{Regge limit of the $\cN=4$ four-point amplitude}
\setcounter{equation}{0}
\label{sec:oneandtwoloop}

In this section, we briefly review 
the Regge limit of the $\cN=4$ four-point amplitude 
at tree and one-loop level to establish conventions 
and prepare for the generalization to higher loops
in subsequent sections.
\para

The tree-level four-gluon amplitude is 
\be
\cA^\Zero = \sum_{\lam=1}^{3} A_\lam^\Zero t_\lam^\Zero
\ee

with color-ordered amplitudes\footnote{Anticipating the need 
to dimensionally regularize loop amplitudes, 
we take the spacetime dimension as $D=4-2 \de$.
Choosing $g$ to be a dimensionless parameter,
the coupling constant in $D$ dimensions is 
$g_D = g \mu^\de$, where $\mu$ is a renormalization scale.
\label{gconvention}
}
\be
A_1^\Zero = - g^2 \mu^{2\de} {4  K \over s t }, \qquad
A_2^\Zero = - g^2 \mu^{2\de} {4 K \over s u }, \qquad
A_3^\Zero = - g^2 \mu^{2\de} {4 K \over t u } \,.
\label{treelevelcolorordered}
\ee

The polarization-dependent factor $K$
is defined in eq.~(7.4.42) of ref.~\cite{Green:2012oqa}
and the Mandelstam variables are 
\begin{align}
s = (k_1 + k_2)^2, \qquad
t = (k_1 + k_4)^2, \qquad
u = (k_1 + k_3)^2 \,.
\end{align}
The components in \eqn{treelevelcolorordered} manifestly obey 
\eqn{treelevelgrouptheory} by virtue of
massless momentum conservation $s+t+u=0$.
\para

In the Regge limit $s\gg -t$ (so that $u \approx -s$),
one has $A_2^\Zero \ll A_1^\Zero$ and $A_3^\Zero \approx -A_1^\Zero$
so that 
\be
\cA^\Zero 
\quad \mathrel{\mathop{\longrightarrow}\limits_{s \gg -t}}\quad
A_1^\Zero \CX_\one 
\label{fulltreelevel}
\ee
where we define 
\be
\CX_\one  \equiv t_1^\Zero - t_3^\Zero \,.
\label{C00def}
\ee
It is straightforward (using the conventions of footnote \ref{Tconvention})
to show that $\CX_\one$ corresponds to the
tree-level $t$-channel color factor
\be
\CX_\one = \tf^{a_4 a_1 e} \tf^{a_2 a_3 e} 
\label{ff}
\ee
where $\tf^{abc} \equiv  i \sqrt2 f^{abc}$.
\para

We digress slightly to discuss the behavior 
of the tree-level amplitude under crossing symmetry
which exchanges external legs 2 and 3.
From their definitions,
we can see that the trace basis color factors
(\ref{extendedbasis}) obey
\be
t^\Ell_{1+3j} \leftrightarrow t^\Ell_{3+3j}, 
\qquad 
t^\Ell_{2 + 3j} \to t^\Ell_{2 + 3j}
\label{extendedexchange}
\ee
under $a_2 \leftrightarrow a_3$
and thus the color factor (\ref{C00def}) is odd:
$C_\one \to -C_\one$.
In the Regge limit, $u \approx -s$, 
so the Mandelstam variables transform 
under $k_2 \leftrightarrow k_3$ as 
\be 
s \to  -s, \qquad t \to t
\ee
while the factor $K$ is invariant under all permutations of
external legs.
Thus the tree-level color-ordered amplitude $A_1^\Zero$
is also odd:  $A_1^\Zero \to -  A_1^\Zero$.
The full tree-level amplitude (\ref{fulltreelevel}),
being the product of two odd objects, is thus invariant,
as expected from Bose symmetry.
\para

Next, we turn to the one-loop 
$\cN=4$ SYM four-gluon amplitude \cite{Bern:1998ug}
\be
\cA^\One = - i g^4 \mu^{2\de} (4 K )
\left[   
  \cI^\One (s,t) \ c^\One_{1234}
+ \cI^\One (u,s) \ c^\One_{1243}
+ \cI^\One (t,u) \ c^\One_{1324}
\right]
\label{oneloopamp}
\ee

where  \cite{Bern:1993kr}
\begin{align}
\cI^\One (s,t)
&= 
 \mu^{2\de} 
\int {d^{4-2\de}  p \over (2 \pi)^{4-2\de} } 
{1 \over p^2 (p-k_1)^2 (p-k_1-k_2)^2 (p+k_4)^2 } 
\nn\\[2mm]
&= 
{i \over 8 \pi^2}  {(4 \pi \e^{-\gamma} )^\de r(\de) \over  s t \de^2 }
\left[ 
   \left( \mu^2\over -s \right)^{\de} F\left(\de, 1+{s\over t}\right)
+ \left( \mu^2 \over -t \right)^{\de} F\left(\de, 1+{t \over s} \right) \right]
\label{oneloopintegral}
\end{align}

with  
\begin{align}
F(\de,z)  
&\equiv
{}_2F_1 (1, -\de, 1-\de, z) \,,
\nn\\[2mm]
r(\de)
&\equiv
\e^{\gamma \eps} {\Gamma(1+\eps) \Gamma^2(1-\eps) \over \Gamma(1-2\eps)} 
= 1 - \fr{1}{2} \zeta_2  \eps^2
- \fr{7}{3} \zeta_3  \eps^3 - \fr{47}{16} \zeta_4  \eps^4 + \cdots
\,.
\end{align}

The one-loop color factors 
$c^\One_{1234} \equiv
 \tf^{e a_1 b} \tf^{b a_2 c} \tf^{c a_3 d} \tf^{d a_4 e}$, 
etc.  
can be written in the trace basis as 
\begin{align} 
c^\One_{1234}
& 
=  t_1^\One + 2 (t_4^\One + t_5^\One + t_6^\One) \,,
\nn\\
c^\One_{1243}
& 
=  t_2^\One + 2 (t_4^\One + t_5^\One + t_6^\One) \,,
\nn\\
c^\One_{1324}
&
=  t_3^\One + 2 (t_4^\One + t_5^\One + t_6^\One) \,.
\label{oneloopcolorfactors}
\end{align}

Inserting these into \eqn{oneloopamp},
we obtain one-loop color-ordered amplitudes
\begin{align}
A_1^\One
&
\,=\,  A_1^\Zero  \atil  \, {1 \over  \de^2 } 
\left[ 
-\left(  \e^{i \pi }  x \right)^{\de} F\left(\de, 1-{1\over x}\right)
- F\left(\de, 1-x  \right) 
\right] \,,
\nn\\[2mm]
A^\One_2 
&
\,=\,   A_1^\Zero  \atil   \, {1 \over  \de^2 } \, 
\left({x \over 1-x}\right)
\left[ 
-\left(  \e^{i \pi }  x \right)^{\de} F\left(\de, {-x \over 1-x}\right)
- \left( x \over 1-x \right)^{\de} F\left(\de, x  \right) 
\right] \,,
\nn\\[2mm]
A^\One_3 
&\,=\,
  A_1^\Zero  \atil \,   {1 \over  \de^2 }  \, 
\left( {1 \over 1-x}  \right)
\left[ 
F\left(\de, {1\over 1-x}\right)
+ \left( x \over 1-x \right)^{\de} F\left(\de, {1 \over x} \right) 
\right] \,,
\nn\\[2mm]
A_4^\One 
&\,=\,A_5^\One \,=\,A_6^\One 
 \,=\,  2 (A_1^\One +A_2^\One +A_3^\One ) \,,
\end{align}

where we define the dimensionless parameters
\begin{align}
\atil 
&\equiv  
{g^2  \over 8 \pi^2} 
{ \Gamma^2(1-\eps) \Gamma(1+\eps) \over \Gamma(1-2\eps)} 
\left( 4 \pi \mu^2  \over -t \right)^{\de}  \,,
\nn\\
x
&\equiv
{-t \over s} \,.
\end{align}
We analytically continue to positive $s$ using $-s = \e^{-i\pi} s$,
so that $(-t)/(-s) = \e^{i\pi} x$.
\para

To take the Regge limit,
one expands the hypergeometric functions in $x$, 
keeping only the $\cO(x^0)$ term 
\begin{align}
A_1^\One
&
\,=\,  A_1^\Zero  \atil   
\left[ \, -\,  {2 \over \de^2}\,  -\,  {\log x + i \pi\over \de} \,  +\,  f_1 (\de) \right]
+   \cO( x) 
\,,
\nn\\[2mm]
A^\One_2 
&
\,=\,   \cO( x) 
\,,
\nn\\[2mm]
A^\One_3 
&\,=\,
  A_1^\Zero  \atil   
\left[   {2 \over \de^2}\,  +\,  {\log x\over \de} \, -\,  f_1 (\de) \right]
+   \cO( x) 
\,,
\nn\\[2mm]
A^\One_4 
&
\,=\,A^\One_5 
\,=\,A^\One_6 
\,=\,  A_1^\Zero  \atil   \left[{- 2 \pi i  \over  \de } \right]
+   \cO( x) 
\label{oneloopcolororderedregge}
\end{align}

where 
\begin{align}
f_1(\de) 
&\equiv
{2 \over \de^2} - \psi(-\de) -  \pi  \cot (\pi \de )  - \gamma 
\nn\\
&
=
\sum_{m=0}^\infty [ 2 + (-1)^m ] \zeta_{m+2}  \eps^{m}   \,.
\end{align}

Two- and higher-loop amplitudes for $\cN=4$ four-gluon amplitudes
are known in terms of linear combinations of planar and nonplanar 
integrals \cite{Bern:1997nh,Bern:1998ug,Bern:2007hh,Bern:2008pv,Bern:2010tq,
Bern:2012uf,Bern:2012uc}.
The two-loop planar integrals were evaluated 
in Laurent expansions through $\cO(\de^2)$,
and the three-loop planar integrals through $\cO(\de^0)$,
in ref.~\cite{Bern:2005iz}.
More recently, the Laurent expansions
of the corresponding nonplanar integrals
have been evaluated to the same accuracy 
\cite{Tausk:1999vh,Henn:2013tua,Henn:2013nsa,Henn:2020lye}.
Using these, the color-ordered four-point amplitudes through 
three loops have been evaluated in ref.~\cite{Henn:2016jdu}
with results given in ancillary files.
We have used these results to derive the Regge limits of these
amplitudes.
As at tree level and one loop, the color-ordered amplitude $A_2^\Ell$
is suppressed relative to the other color-ordered amplitudes, 
and this holds to all 
orders.\footnote{This is because $A_2^\Ell$ 
is leading in the large-$N$ (planar) limit, and the
BDS ansatz \cite{Bern:2005iz} expresses the all-orders planar amplitude
in terms of the tree-level and one-loop amplitude.}
Once we have developed a suitable basis of
higher-loop color factors in sec.~\ref{sec:reggebasis},
and discussed the infrared divergent structure 
of the higher-loop amplitudes in sec.~\ref{sec:IR},
we will be able to present our results in a compact way.

\section{Regge basis of color factors}
\setcounter{equation}{0}
\label{sec:reggebasis}

As noted in sec.~\ref{sec:constraints},
the $\ell$-loop color-ordered amplitudes $A_\lam^\Ell$,
coefficients in the extended trace basis $t_\lam^\Ell$
with $\lam = 1, \cdots, 3\ell + 3$,
are not independent due to group-theory constraints.
Moreover, $A_2^\Ell$ is suppressed 
in the Regge limit relative to the other color-ordered amplitudes, 
so taking into account the group-theory constraints,
there remain {\it in the Regge limit}
one independent amplitude at tree level,
two independent amplitudes at one loop,
and $3\ell-2$ independent amplitudes for $\ell \ge 2$.
\para

In this section, we develop an alternative basis $\CX_\ik$ 
of color factors that will be useful for characterizing 
independent amplitudes in the Regge limit.
In particular,
the color factors $\CX_\ik$
are chosen to have definite signature 
under the exchange of external legs 2 and 3.
We then express the coefficients $\BX^\Ell_\ik$ 
of the $\ell$-loop amplitude in this basis 
as linear combinations of the color-ordered amplitudes $A_\lam^\Ell$.
\para

First, let's count the number of 
independent even and odd signature amplitudes. 
As noted in sec.~\ref{sec:oneandtwoloop},
the exchange of legs 2 and 3
acts on the $3\ell+3$ elements 
of the extended trace basis (\ref{extendedbasis})
according to \eqn{extendedexchange},
so one can form linear combinations 
with $2\ell+2$ even elements
$t^\Ell_{1+3j} + t^\Ell_{3+3j}$ and $\t_{2+3j}^\Ell$,
and $\ell+1$ odd elements
$t^\Ell_{1+3j} - t^\Ell_{3+3j}$.
The null vectors that correspond to group-theory constraints 
can also be chosen to have definite signature. 
The tree-level null vector (\ref{treelevelnull}) is even,
there are two even and one odd one-loop null vectors (\ref{oneloopnull}),
and for two loops and above, there are three even and one
odd null vectors, cf. \eqns{twoloopnull}{threeloopnull}.
In the Regge limit, the coefficient of $t_2^\Ell$ is suppressed,
eliminating one more even element.
The upshot is that there remain
{\it in the Regge limit} 
one odd independent amplitude at tree level,
one odd and one even independent amplitude at one loop,
and 
$\ell$ odd 
and 
$2\ell-2$ even 
independent amplitudes
for $\ell \ge 2$.   
Our choice of basis must reflect this.
\para

The Regge basis of color factors $\CX_\ik$ will be built from 
operators acting on the odd tree-level color factor
$\CX_\one$ given by \eqn{C00def}.
In the color-space formalism introduced in 
refs.~\cite{Catani:1996jh,Catani:1996vz,Catani:1998bh},
the color operator $\bT_i$ 
acts on a color factor by inserting a generator 
in the adjoint representation of SU($N$) 
on the $i$th external leg,
specifically $(\bT^a_i)^{bc} = {1 \over \sqrt2} \tf^{bac}$,
while acting as the identity on all the other external legs.
Color conservation implies that 
\be 
\sum_{i=1}^4 \bT_i = 0
\label{colorconservation}
\ee
when acting on a four-point amplitude \cite{Catani:1998bh}.
\para

Applying the operator
$\bT_i \cdot \bT_j \equiv \sum_a \bT_i^a \bT_j^a$
to a given color factor 
acts to attach a rung between external legs $i$ and $j$.
For example, the operators $\bT_1 \cdot \bT_2 $ and $\bT_1 \cdot \bT_3 $
convert the tree-level $t$-channel color factor $\CX_\one$ 
to a one-loop box and crossed box color factor, respectively,
but  $\CX_\one$ is an eigenstate of $\bT_1 \cdot \bT_4 $,
since  $\tf^{dae} \tf^{ebg} \tf^{gcd} = N \tf^{abc}$.
More precisely,
\be
\bT_1 \cdot \bT_2 
\CX_\one 
=  - \half c_{1234}^\One
\,, \qquad
\bT_1 \cdot \bT_3 
\CX_\one 
=
\half c_{1342}^\One
\,, \qquad
\bT_1 \cdot \bT_4 
\CX_\one 
=
- \half N \CX_\one  \,.
\label{addingrungs}
\ee

Also 
\be
(\bT_i^2)^{bc}  =  \half \tf^{bad}\tf^{dac} = N \delta^{bc}
\ee
for any $i$,
so that \eqn{colorconservation} implies 
\be
 \bT_3 \cdot \bT_4 = \bT_1 \cdot \bT_2, \qquad
 \bT_2 \cdot \bT_4 = \bT_1 \cdot \bT_3 , \qquad
 \bT_2 \cdot \bT_3 = \bT_1 \cdot \bT_4  \,.
\ee

Next, following
refs.~\cite{Dokshitzer:2005ig,Bret:2011xm,DelDuca:2011ae},
we define color operators associated with color flow in each channel
\be
\bT_s = \bT_1 + \bT_2, \qquad
\bT_u =\bT_1 + \bT_3, \qquad
\bT_t =\bT_1 + \bT_4
\ee
with 
color conservation (\ref{colorconservation})
implying $\bT_s^2 + \bT_t^2 + \bT_u^2 = 4N$.
Finally we define \cite{Caron-Huot:2017fxr}
\be
\bT_{s-u}^2 \equiv \half \left( \bT_s^2 - \bT_u^2 \right) \,.
\ee

A key observation is that the operators
\begin{align}
\bT_t^2 &=  2N + 2\bT_1 \cdot \bT_4
= - 2 ( \bT_1\cdot \bT_2 +  \bT_1\cdot \bT_3 ), \nn\\[2mm]
\bT_{s-u}^2 &=  \bT_1\cdot \bT_2 -  \bT_1\cdot \bT_3 
\label{keyobs}
\end{align}

are even and odd, respectively, under the exchange of external 
legs $2$ and $3$ (which implies $s \leftrightarrow u$ and $t\to t$).
\para

\subsection{One-loop basis}

At one loop, we have one independent odd amplitude
and one independent even amplitude.
Recalling that both $\CX_\one$ and $\bT_{s-u}^2$ are odd 
under $2\leftrightarrow 3$, and defining
\be
\CX_\two \equiv \bT_{s-u}^2 \CX_\one
\ee
we form a one-loop basis of color factors
$\{ N \CX_\one, \CX_\two \}$ whose elements have signature $\{ - , + \}$.
Their components in the one-loop trace basis $t_\lam^\One$ 
\begin{align}
\begin{array}{rrrrrr}
N \CX_\one = 
(1,& 0,&-1,&  0,&  0,&  0), \\[2mm]
\CX_\two 
=   (-\half,& 0,& -\half,& -2,& -2,& -2), \\
\end{array}
\end{align}

can be obtained by using \eqns{oneloopcolorfactors}{addingrungs}.
Expressing the amplitude in this one-loop basis
\begin{align}
\cA^\One 
&=
A_1^\Zero
\atil 
\left[   
\BX_\one^\One  N \CX_\one
+
 \BX_\two^\One  \CX_\two
\right] 
\label{oneloopreggebasis}
\end{align}

we obtain 
\begin{align}
A_1^\Zero \atil ~ \BX_\one^\One
&= 
\fr{1}{2} (  A_1^\One - A_3^\One ) \,,
\nn\\[2mm]
A_1^\Zero \atil ~\BX_\two^\One
&= 
- ( A_1^\One + A_3^\One )  \,.
\label{oneloopreggetrace}
\end{align}

The coefficients $\BX_\one^\One$ and $\BX_\two^\One$
inherit the same signature under crossing symmetry
(which takes $A_1^\One \leftrightarrow A_3^\One$)
as their associated color factors,
leaving the whole amplitude Bose symmetric.
\para

Using \eqn{oneloopcolororderedregge} in \eqn{oneloopreggetrace}, 
we obtain
\begin{align}
\BX_\one^\One &=
 - {2 \over \de^2} + {L \over \de}   +f_1(\de) \,,
\nn\\
\BX_\two^\One &=
{i \pi \over \de}
\label{oneloopreggecoefficients}
\end{align}

where  we define \cite{Caron-Huot:2017fxr}
\be
L \equiv  - \log x - \half i \pi \,.
\label{Ldef}
\ee
Note that $L$ is the Regge limit of
the even-signature combination of logarithms 
\be
- \fr{1}{2} 
\left[ \log \left( -t \over -s \right) 
      + \log \left( -t \over -u \right) \right]
\nn\\
=
- \fr{1}{2} 
\left[ \log \left( \e^{i \pi} x  \right) 
      + \log \left( x \over 1-x  \right) \right]
\quad\mathrel{\mathop{\longrightarrow}\limits_{x \to 0 }} \quad
L \,.
\ee
As noted in ref.~\cite{Caron-Huot:2017fxr},
when expressed in terms of the natural variable $L$,
the coefficients of odd-signature color factors are real, 
whereas the coefficients of even-signature color factors 
are imaginary.
We will see this at higher loops as well.

\subsection{Two-loop basis}

At two loops, we have two independent odd amplitudes
and two independent even amplitudes.
Thus, to the previous two color factors, we add two more
\begin{align}
\CX_\three &=   [\bT_t^2, \bT_{s-u}^2] \CX_\one \,, \nn\\[2mm]
\CX_\four &=   ( \bT_{s-u}^2)^2 \CX_\one \,, 
\end{align}
to form a two-loop basis of color factors 
$\{ N^2 \CX_\one, N \CX_\two, \CX_\three, \CX_\four \}$ 
whose elements have signature $\{ - , +, +, - \}$.
In order to obtain explicit expressions for these color factors,
it is convenient to express the operators 
$\bT_t^2$ and $\bT_{s-u}^2$ as matrices\footnote{
The matrices given here,
which also appear in appendix C of ref.~\cite{Caron-Huot:2017fxr},
are the transpose of those 
in ref.~\cite{Naculich:2011ep}.  
This is because here we take these matrices 
to act to the right on the color factors,
whereas in ref.~\cite{Naculich:2011ep} the matrices acted to the left.}
\be
\bT_t^2 = 
\left(
\begin{array}{rrrrrr}
 N & 0 & 0 & 0 & 0 & -1 \\ [1mm]
 0 & 2 N & 0 & 1 & 0 & 1 \\ [1mm]
 0 & 0 & N & -1 & 0 & 0 \\ [1mm]
 0 & 2 & 0 & 2 N & 0 & 0 \\ [1mm]
 -2 & 0 & -2 & 0 & 0 & 0 \\ [1mm]
 0 & 2 & 0 & 0 & 0 & 2 N \\
\end{array}
\right)\,,
\qquad
\bT_{s-u}^2 = 
\left(
\begin{array}{rrrrrr}
 -\frac{N}{2} & 0 & 0 & 0 & -1 & -\frac{1}{2} \\[1mm]
 0 & 0 & 0 & -\frac{1}{2} & 0 & \frac{1}{2} \\ [1mm]
 0 & 0 & \frac{N}{2} & \frac{1}{2} & 1 & 0 \\ [1mm]
 0 & 1 & 2 & N & 0 & 0 \\ [1mm]
 -1 & 0 & 1 & 0 & 0 & 0 \\ [1mm]
 -2 & -1 & 0 & 0 & 0 & -N \\
\end{array}
\right)
\label{matrices}
\ee

in the original six-dimensional trace basis (\ref{sixdimbasis}).
Using these, 
we can obtain the components 
of the elements of the two-loop basis 
in the extended trace basis $t_\lam^\Two$
\be
\begin{array}{rrrrrrrrrrr}
N^2 \CX_\one =&  
(  1, & 0, & -1, &  0, &  0, &  0, & 0, & 0, & 0), \\ [2mm]
N \CX_\two =&  
( -\half , & 0, & -\half , &  -2, &  -2, &  -2, & 0, & 0, & 0), \\ [2mm]
\CX_\three =&  
(  0, & 0, & 0,&  -2, & 4, & -2, &   2, & -4, &  2), \\ [2mm]
\CX_\four =&  
(  \fr{1}{4} & 0, & -\fr{1}{4},  &  -3, &  0,& 3, & 3, & 0, & -3). \\
\end{array}
\ee

Expressing the amplitude in this two-loop basis 
\be
\cA^\Two 
= A_1^\Zero \atil^2 
\Big[ 
 \BX_\one^\Two N^2    \CX_\one
+ \BX_\two^\Two  N \CX_\two
+ \BX_\three^\Two \CX_\three
+ \BX_\four^\Two \CX_\four
\Big]
\label{twoloopreggebasis}
\ee

we obtain\footnote{It is easiest to obtain these results
(and similarly at higher loops)
by expressing $t_\lam^\Two$ in terms of 
an enlarged basis consisting of 
$\{ N^2 \CX_\one, N \CX_\two, \CX_\three, \CX_\four \}$ 
together with $t_2^\Two$ and the two-loop null vectors 
(\ref{twoloopnull}).
The coefficients of $t_2^\Two$ and the null vectors 
will automatically vanish by virtue of $A_2^\Two=0$
and the group theory constraints (\ref{twoloopgrouptheory}).}
\begin{align}
A_1^\Zero \atil^2 ~\BX_\one^\Two 
&= 
\fr{1}{2} (  A^\Two_1 - A^\Two_3 )
- \fr{1}{24} ( A^\Two_7 - A^\Two_9) \,,
\nn\\[2mm]
A_1^\Zero \atil^2 ~\BX_\two^\Two 
&= 
- ( A^\Two_1 + A^\Two_3 )  \,,
\nn\\[2mm]
A_1^\Zero \atil^2 ~\BX_\three^\Two 
&= 
\fr{1}{4} ( A^\Two_7 + A^\Two_9 ) \,,
\nn\\[2mm]
A_1^\Zero \atil^2 ~\BX_\four^\Two 
&= 
\fr{1}{6} ( A^\Two_7 - A^\Two_9 ) \,.
\label{twoloopreggetrace}
\end{align}
As we saw previously, 
the coefficients $\BX_\ik^\Two$ 
inherit the same signature under crossing symmetry
(which takes $A_1^\Two \leftrightarrow A_3^\Two$,
and $A_7^\Two \leftrightarrow A_9^\Two$)
as their associated color factors,
leaving the whole amplitude Bose symmetric.
We will use \eqn{twoloopreggetrace} in sec.~\ref{sec:IR}
to compute the Laurent expansions of $\BX_\ik^\Two$
through $\cO(\de^2)$.

\subsection{Three-loop basis}

At three loops, 
we have three independent odd amplitudes
and four independent even amplitudes.
To the previous four color factors, we add three more
\begin{align}
\CX_\five
&=   [\bT_t^2,  [\bT_t^2,  \bT_{s-u}^2]] \CX_\one \,, \nn\\[2mm]
\CX_\six
&=   [\bT_{s-u}^2, [\bT_t^2,  \bT_{s-u}^2]] \CX_\one \,, \nn\\[2mm]
\CX_\seven
&=   (\bT_{s-u})^3 \CX_\one \,, 
\end{align}

to form a three-loop basis of color factors
$\{ N^3 \CX_\one, N^2 \CX_\two, N\CX_\three, N \CX_\four,
\CX_\five, \CX_\six, \CX_\seven \}$ 
whose elements have 
signature $\{ - , +, +, -, +, -, + \}$.
The elements of the three-loop basis 
have components in the extended trace basis $t_\lam^\Three$  given by
\be
\begin{array}{rrrrrrrrrrrrrr}
N^3 \CX_\one =&  
(  1, & 0, & -1, &  0, &  0, &  0, & 0, & 0, & 0,& 0,& 0,& 0), \\ [2mm]
N^2 \CX_\two =&  
( -\half , & 0, & -\half , & -2,& -2, & -2, & 0, & 0, & 0,& 0,& 0,& 0), \\ [2mm]
N \CX_\three =&  
(  0, & 0, & 0,&  -2, & 4, & -2, &   2, & -4, &  2,& 0,& 0,& 0), \\ [2mm]
N \CX_\four =&  
(\fr{1}{4}, & 0, & -\fr{1}{4},&-3,&0,& 3,&3,& 0,& -3,& 0,& 0,& 0), \\ [2mm]
\CX_\five =&  
(  0, &  0, &  0, &- 2, &-  4, &-  2, &2, &- 8, &2, &- 8, &- 8, &- 8), \\ [2mm]
\CX_\six
=&  
(  0, &  0, &  0, &- 1, &  0, &1, &-  5, &  0, &5, &  0, &  0, &  0), \\ [2mm]
 \CX_\seven
=&  ( - \fr{1}{8} &  0, &- \fr{1}{8},  &-  \fr{7}{2}, &- \fr{1}{2}, 
&-  \fr{7}{2}, &- 3, &  3, &- 3, &- 6, &- 6, &- 6). \\
\end{array}
\label{threeloopcomponents}
\ee

Expressing the amplitude in this three-loop basis 
\be
\cA^\Three
= A_1^\Zero \atil^3
\Big[ 
\BX_\one^\Three N^3   \CX_\one
+ \BX_\two^\Three N^2    \CX_\two
+ \BX_\three^\Three N     \CX_\three
+ \BX_\four^\Three N    \CX_\four
+ \BX_\five^\Three \CX_\five
+ \BX_\six^\Three   \CX_\six
+ \BX_\seven^\Three    \CX_\seven
\Big]
\label{threeloopreggebasis}
\ee

we may derive
\begin{align}
A_1^\Zero \atil^3 ~\BX_\one^\Three 
&=
\fr{1}{2} (A^\Three_1-A^\Three_3) +\fr{5}{144} (A^\Three_4- A^\Three_6)  -\fr{1}{144} (A^\Three_7-A^\Three_9)  \,,
\nn\\[2mm]
A_1^\Zero \atil^3 ~\BX_\two^\Three 
&=
-\fr{13 }{12}(A^\Three_1 + A^\Three_3) +\fr{1}{48}(A^\Three_4+A^\Three_6) +\fr{1}{48}(A^\Three_7+A^\Three_9)  \,,
\nn\\[2mm]
A_1^\Zero \atil^3 ~\BX_\three^\Three 
&=
\fr{3}{4}(A^\Three_1 +   A^\Three_3)   
-\fr{3}{16}(A^\Three_4 + A^\Three_6)
+\fr{5}{16} (A^\Three_7+ A^\Three_9) +\fr{1}{4}A^\Three_8 \,,
\nn\\[2mm]
A_1^\Zero \atil^3 ~\BX_\four^\Three 
&=
-\fr{5}{36}(A^\Three_4- A^\Three_6) +\fr{1}{36}(A^\Three_7-A^\Three_9) \,,
\nn\\[2mm]
A_1^\Zero \atil^3 ~\BX_\five^\Three 
&=
-\fr{1}{4}(A^\Three_1+A^\Three_3)
+\fr{1}{16}(A^\Three_4 + A^\Three_6) 
-\fr{3}{16}(A^\Three_7+ A^\Three_9) -\fr{1}{4}A^\Three_8 \,,
\nn\\[2mm]
A_1^\Zero \atil^3 ~\BX_\six^\Three 
&=
-\fr{1}{12}(A^\Three_4-A^\Three_6) -\fr{1}{12}(A^\Three_7-A^\Three_9) \,,
\nn\\[2mm]
A_1^\Zero \atil^3 ~\BX_\seven^\Three 
&=
\fr{1}{3}(A^\Three_1 + A^\Three_3) -\fr{1}{12}(A^\Three_4 +A^\Three_6) -\fr{1}{12}(A^\Three_7 + A^\Three_9) \,.
\label{threeloopreggetrace}
\end{align}

We will use these expressions in sec.~\ref{sec:IR}
to compute the Laurent expansions of $\BX_\ik^\Three$
through $\cO(\de^0)$.

\subsection{Higher-loop basis}

For each additional loop, the number of 
independent odd amplitudes increases by one   
and the number of 
independent even amplitudes increases by two.
Thus to obtain the $\ell$-loop basis for independent amplitudes
we add to  the $(\ell-1)$-loop basis 
three additional color factors
\begin{align}
\CX_{\ell 1} 
&=   [\bT_t^2, \cdots, [\bT_t^2,  [\bT_t^2,  \bT_{s-u}^2]]\cdots ] \CX_\one \,, \nn\\[2mm]
\CX_{\ell,\ell-1} 
&=   [\bT_{s-u}^2, \cdots, [\bT_{s-u}^2, [\bT_t^2,  \bT_{s-u}^2]]\cdots ] \CX_\one \,, \nn\\[2mm]
\CX_{\ell\ell} 
&=   (\bT_{s-u})^\ell \CX_\one \,, 
\end{align}
where each color factor contains exactly $\ell$ operators $\bT^2$.
The operator $\CX_\ellk$ contains $k$ factors of $\bT_{s-u}^2$
and thus has odd signature if $k$ is even, and even signature if $k$ is odd.
Thus $\CX_{\ell 1} $ has even signature,
and 
$\CX_{\ell,\ell-1} $
and 
$\CX_{\ell\ell} $ have opposite signatures, as required.
This choice of basis is motivated by the Regge limit 
of the structure of IR divergences as studied in 
refs.~\cite{DelDuca:2011ae,Caron-Huot:2013fea,DelDuca:2014cya,Caron-Huot:2017fxr}.
\para

The $\ell$-loop amplitude can then be expressed in this 
basis as 
\be
\cA^\Ell
= A_1^\Zero \atil^\ell
\sum_{i=0}^\ell \sum_{k}
\BX_\ik^\Ell N^{\ell-i}   \CX_\ik
\label{reggeexpansion}
\ee
where the range of $k$ is
\be
k = 
\begin{cases}
0, & \mbox{when } i=0, \\
1, & \mbox{when } i=1,  \\
1, 2, & \mbox{when } i=2, \\
1, i-1, i,  & \mbox{when } i \ge 3. 
\end{cases}
\ee
Using \eqn{matrices}, the $\ell$-loop Regge color factors $\CX_\ellk$ 
can be expressed in terms of  $t_\lam^\Ell$, and
the coefficients $\BX_\ik^\Ell$ obtained as linear combinations 
of the 
$\ell$-loop color-ordered amplitudes $A_\lam^\Ell$.
\para

The expressions for $\BX_\ik^\Ell$ grow
increasingly complicated at higher loops, 
but the general expression for one of them, $\BXL$,
can be easily guessed from explicit results (obtained through $\ell=9$).
The expressions differ depending on whether $\ell$ is even or odd.
In the former case, 
\begin{align}
A_1^\Zero \atil^\ell \BXL 
&= \frac{1}{2 \cdot 3^{\ell/2} } 
\left( A_{3\ell+1}^\Ell - A_{3 \ell+3}^\Ell\right)
\nn\\[2mm]
&= \frac{1}{2 \cdot 3^{\ell/2} } 
\left( A_1^{(\ell,\ell)} -A_3^{(\ell,\ell)}  \right), 
\qquad \hbox{for even } \ell \,.
\label{evenBXL}
\end{align}

In the latter case,  we have 
\begin{align}
A_1^\Zero \atil ~\BX_\two^\One
&= 
- ( A_1^\One + A_3^\One )  
\nn\\
&= - ( A_1^{(1,0)} + A_3^{(1,0)})
\end{align}
for $\ell=1$, and for odd $\ell > 1$, we have 
\begin{align}
A_1^\Zero \atil^\ell \BXL 
& =
\frac{1}{   3^{(\ell-1)/2} } 
\left[  \left( A_{3\ell-8}^\Ell + A_{3 \ell-6}^\Ell \right)
- \fr{1}{4} \left( A_{3\ell-5}^\Ell + A_{3 \ell-3}^\Ell 
+    A_{3\ell-2}^\Ell + A_{3 \ell}^\Ell\right)
\right] 
\label{oddBXL}
\\[2mm]
&= 
\frac{1}{   3^{(\ell-1)/2} } 
\left[ \left( A_1^{(\ell,\ell-3)} +A_3^{(\ell,\ell-3)} \right)
-\fr{1}{4}  \left( A_4^{(\ell,\ell-2)} +A_6^{(\ell,\ell-2)} 
 + A_1^{(\ell,\ell-1)} +A_3^{(\ell,\ell-1)} \right)
\right], 
\nn\\[2mm]
& \hskip90mm 
\qquad \hbox{for odd } \ell > 1
\nn
\end{align}

where we have expressed these in terms of color-ordered amplitudes in 
both the $(3\ell+3)$-dimensional extended trace basis (\ref{decomp})
and also the original six-dimensional trace basis (\ref{expandeddecomp}).
\para

In refs.~\cite{Naculich:2008ys,Naculich:2009cv}, we showed 
that the leading IR divergence of 
the color-ordered amplitude $A^\Ellk$ is $1/\de^{2\ell-k}$,
with planar amplitudes 
$A^{(\ell,0)}_\lam$ 
having the most severe $1/\de^{2\ell}$ divergences,
and the most-subleading-color amplitudes $A^{(\ell,\ell)}_\lam$
having at most a $1/\de^\ell$ divergence.
At the end of sec.~\ref{sec:IR}, 
we will show that $\BXL$ also has at most 
a $1/\de^\ell$ divergence.
For even $\ell$, this is manifest from \eqn{evenBXL}, 
where $\BXL$ is expressed in terms of color-ordered amplitudes 
that are most-subleading in the $1/N$ expansion.
For odd $\ell$, however, \eqn{oddBXL} shows that this is not the case,
so the $1/\de^\ell$ behavior of $\BXL$ requires some
intricate cancellations of the more severe IR divergences
appearing in 
$A^{(\ell,\ell-3)}_\lam$, $A^{(\ell,\ell-2)}_\lam$,  
and $A^{(\ell,\ell-1)}_\lam$.

\section{IR-divergence structure of the $\cN=4$ SYM amplitude} 
\setcounter{equation}{0}
\label{sec:IR}

In this section we first briefly review 
the known structure of infrared divergences 
of the $\cN=4$ SYM four-point amplitude through 
three loops \cite{Catani:1998bh,Sterman:2002qn,Bern:2005iz,
Aybat:2006wq,Aybat:2006mz,
Dixon:2008gr,Becher:2009cu,Gardi:2009qi,Dixon:2009gx,
Becher:2009qa,Gardi:2009zv,Dixon:2009ur,Almelid:2015jia,
Bret:2011xm,DelDuca:2011ae,Caron-Huot:2013fea,
DelDuca:2013ara,DelDuca:2014cya,Caron-Huot:2017fxr},
focusing in particular on the Regge limit.
We then take the Regge limit of known results 
for the four-point amplitude at one, two, and three loops \cite{Henn:2016jdu}
to confirm the expected IR divergences
and to extract the IR-finite part of the amplitude in this limit,
writing the result in terms of the Regge basis 
of color factors introduced in sec.~\ref{sec:reggebasis}. 
Finally we compare these to results
obtained via an effective Hamiltonian approach 
based on Balitsky-Fadin-Kuraev-Lipatov theory 
in refs.~\cite{Caron-Huot:2013fea,Caron-Huot:2017fxr}.
\para  

The amplitude
may be factored into jet, soft, and hard 
functions \cite{Sterman:2002qn,Aybat:2006wq,Aybat:2006mz}
\be
\cA \left( {s_{ij}\over \mu^2} \right)
=  J \left( {Q^2\over \mu^2} \right)
~ \bS \left({s_{ij}\over Q^2},   {Q^2\over \mu^2} \right)
~ \cH \left({s_{ij}\over Q^2},   {Q^2\over \mu^2} \right)
\ee
where the factors $J$ and $\bS$ characterize the long-distance
IR-divergent
behavior, and $\cH$, which is IR-finite, 
characterizes the short-distance behavior.
Here $s_{ij} = (k_i+k_j)^2$, $\mu$ is the renormalization scale, 
and $Q$ is an arbitrary factorization scale that serves
to separate the long- and short-distance behavior.
Since we are interested in the Regge limit $s \gg -t$,
we choose the factorization scale as $Q^2 = -t$
in this paper.
\para

Because $\cN=4$ SYM theory is conformally invariant, 
the jet function may be explicitly evaluated as \cite{Bern:2005iz}
\be
J \left( {-t \over \mu^2} \right)
= 
\exp
\left[ - \suml \atil^\ell  N^\ell 
 \left(
 \frac{\gamma^\Ell }{2( \ell \de)^2}
+ \frac{\cG_0^\Ell } {\ell\de} 
 \right)
\right]
\label{J}
\ee
where  $\gamma^\Ell$ and $\cG_0^\Ell$ are the coefficients of the cusp 
and collinear anomalous dimensions
\be
\gamma^\Ell 
= \{ 4 ,  - 4 \zeta_2 ,  22 \zeta_4 , \cdots  \} \,,
\qquad\qquad 
\cG_0^\Ell
= \{ 0 , - \zeta_3 ,  4 \zeta_5 + \fr{10}{3} \zeta_2 \zeta_3 , \cdots \} \,.
\ee

The soft function $\bS$ depends on the coefficients $\bGam^\Ell$
of the anomalous dimension matrix \cite{Sterman:2002qn}.
The one-loop anomalous dimension matrix is  \cite{Aybat:2006wq,Aybat:2006mz}
\begin{align}
\bGam^\One
& =
\frac{1}{N}  \sum_{i=1}^4 \sum_{j\neq i}^4 \bT_i \cdot \bT_j
 \log \left( Q^2 \over - s_{ij}   \right)
\nn\\
&= 
{4\over N}
\left[
\log \left(-t \over  \e^{-i \pi} s  \right)  \bT_1 \cdot \bT_2 
+ 
\log \left( -t \over -u   \right)  \bT_1 \cdot \bT_3 
\right]
\end{align}

which simplifies in the Regge limit to
\begin{align}
{\bf \Gamma}^\One
&= 
  {4\over N}
\left[
\log \left( \e^{i\pi} x \right)  \bT_1 \cdot \bT_2 
+ 
\log (x) \, \bT_1 \cdot \bT_3 
\right]
\nn\\
&=  {2\over N}
\left[ L \bT_t^2  + i \pi \bT_{s-u}^2 \right] 
\label{oneloopanomdimmatrix}
\end{align}
where in the last line we used \eqns{keyobs}{Ldef}.
On the assumption that the matrices $\bGam^\Ell$ 
commute with one another,
one may explicitly evaluate $\bS$ for $\cN=4$ SYM theory
as\footnote{
The expressions for \eqns{J}{S} differ slightly from those
in ref.~\cite{Naculich:2009cv} 
in that we are using $\atil$ 
rather than 
$a = (g^2 N /8 \pi^2) \left(4 \pi \e^{-\gamma} \right)^\de $
as our loop expansion parameter.
This only affects the form of the infrared-finite hard function $\cH$.}
\cite{Naculich:2009cv}
\be
\bS
=  \exp\left[ \suml 
\atil^\ell
N^\ell
\frac{ \bGam^\Ell}{2\ell\de}\right] \,.
\label{S}
\ee

Commutativity is guaranteed if we assume the anomalous dimension matrix 
is given by the dipole formula \cite{Becher:2009cu,Gardi:2009qi,Dixon:2009gx,
Becher:2009qa,Gardi:2009zv,Dixon:2009ur}
\be
\bGam^\Ell_{\rm dipole} = \fr{1}{4} \gamma^\Ell \bGam^\One  
\label{dipole}
\ee

which is valid through two loops \cite{Aybat:2006wq,Aybat:2006mz},
but receives corrections at three \cite{Almelid:2015jia}
and four \cite{Caron-Huot:2013fea} loops.
If the dipole formula were valid for all loops, 
then using \eqns{oneloopanomdimmatrix}{dipole}
in \eqn{S}, one finds that 
the Regge limit of the soft function can be written in the compact form 
\cite{Bret:2011xm,DelDuca:2011ae}
\be
\bS_{\rm dipole}
= 
\exp \left[ K \left( L \bT^2_t +   i\pi   \bT_{s-u}^2 \right) \right] 
\label{Sdipole}
\ee

where 
\be
K \equiv 
\suml \frac{N^{\ell-1} \gamma^\Ell }{4  \ell \de} \atil^\ell 
= {\atil \over \de} \left( 1 - \half \zeta_2  N \atil
+ \fr{11}{6}  \zeta_4 N^2 \atil^2 + \cdots 
\right)\,.
\label{K}
\ee

The three-loop correction to the dipole formula \cite{Almelid:2015jia}
persists in the Regge limit, and has the effect of modifying 
\eqn{Sdipole} to \cite{Caron-Huot:2017fxr}
\begin{align}
\bS
= 
\exp 
\left[
K \left(
L \bT^2_t +   i\pi   \bT_{s-u}^2
\right)
+ Q_\Delta^\Three
\right] + \cO(\atil^4)
\label{addQ}
\end{align}

where 
\begin{align}
Q_\Delta^\Three
&= \frac{4}{3\de} \atil^3   \Delta^\Three 
\nn\\
\Delta^\Three 
&=
\fr{1}{4} i \pi
\left( \zeta_3 L + 11 \zeta_4 \right)
[\bT_t^2 ,[\bT_t^2 ,\bT_{s-u}^2 ]]
+ 
\fr{1}{4}
\left({\zeta_5-4\zeta_2\zeta_3}  \right)
[\bT_{s-u}^2 ,[\bT_t^2 ,\bT_{s-u}^2 ]]
- 
\fr{1}{4} 
\left( {\zeta_5+2\zeta_2\zeta_3} \right)
\bW
\nn\\
\bW 
&= 
\fr{1}{2}\bigg\{ f^{abe}f^{cde}
\bigg[ \{\bT_t^a, \bT_t^d \}  \Big(\{\bT_{s-u}^b, \bT_{s-u}^c \}
+ \{\bT_{s+u}^b, \bT_{s+u}^c \} \Big)  
\nn\\
&\hskip30mm  +\,   \{\bT_{s-u}^a, \bT_{s-u}^d \}
 \{\bT_{s+u}^b, \bT_{s+u}^c \} \bigg] -\fr{5}{8} N^2 \bT^2_t \bigg\} \,.
\label{Deltathree}
\end{align}

Since the operators in the exponent of \eqn{addQ}  
do not commute with one another, 
it is useful \cite{DelDuca:2011ae} to employ 
a variant of Campbell-Baker-Haussdorf known as the Zassenhaus formula 
\be
\e^{ K(X+Y) }
=
\e^{ KX }
~\e^{ KY }
~\e^{ -(1/2) K^2 [X,Y]  }
~\e^{ (1/6) K^3 \left( [X,[X,Y]] +2[Y,[X,Y]] \right) } + \cO(K^4)
\ee

to write 
\begin{align}
\bS
&
= 
\exp 
\left[ K L \bT^2_t \right]
\exp\left[ i \pi K \bT_{s-u}^2 \right]
\exp\left[ - {i \pi\over 2} L  K^2 [ \bT_t^2, \bT^2_{s-u} ] \right]
\\
&
\hskip5mm
\times 
\exp \left[ 
\frac{i \pi}{6} K^3 L^2 [\bT_t^2 ,[\bT_t^2 ,\bT_{s-u}^2 ]] 
- \frac{\pi^2}{3} K^3 L [\bT_{s-u}^2,[\bT_t^2 ,\bT_{s-u}^2]] 
+ \frac{4}{3\de} \atil^3   \Delta^\Three 
\right] 
+\cO(\atil^4)
\nn
\end{align}

Putting all the pieces together, we write 
the Regge limit of the amplitude 
through $\cO(\atil^3)$ as 
\begin{align}
\cA
&
= 
\exp
\left[ - \suml \atil^\ell  N^\ell 
 \left(
 \frac{\gamma^\Ell }{2( \ell \de)^2}
+ \frac{\cG_0^\Ell } {\ell\de} 
 \right)
\right]
\exp 
\left[ K L \bT^2_t \right]
\exp\left[ i \pi K \bT_{s-u}^2 \right]
\exp\left[ - {i \pi\over 2} L  K^2 [ \bT_t^2, \bT^2_{s-u} ] \right]
\nn\\
&
\hskip5mm
\times 
\exp \left[ 
\frac{i \pi}{6} K^3 L^2 [\bT_t^2 ,[\bT_t^2 ,\bT_{s-u}^2 ]] 
- \frac{\pi^2}{3} K^3 L [\bT_{s-u}^2,[\bT_t^2 ,\bT_{s-u}^2]] 
+ \frac{4}{3\de} \atil^3   \Delta^\Three 
\right]  \cH 
+\cO(\atil^4)
\label{finalamp}
\end{align}
This equation exhibits all of the IR-divergent contributions 
to the amplitude through three loops,
with the IR-finite part encoded in the hard function $\cH$.
To obtain $\cH$, we need to compare \eqn{finalamp} with
known expressions for the amplitude at one, two, and three loops.
This we now proceed to do.

\subsection{Reduced  amplitude}

Expanding \eqn{finalamp} in powers of the loop expansion parameter $\atil$,
it is apparent that the first exponential term in \eqn{finalamp}
is responsible for the most IR-divergent 
terms in the Laurent expansion at $\ell$ loops,
starting with an $\cO(1/\de^{2\ell})$ term.
It is also apparent that the second exponential term in \eqn{finalamp}
is responsible for the leading log behavior,
causing $\cA^\Ell$ to go as $\log^\ell |s/t| $ at $\ell$ loops,
and leading to Reggeization 
\cite{Bret:2011xm,DelDuca:2011ae}.
The amplitude, however, 
also has an intricate structure of subleading logarithms, 
to which the remaining terms in \eqn{finalamp} contribute.
To isolate this subleading logarithmic behavior, 
it is useful \cite{Caron-Huot:2013fea,Caron-Huot:2017fxr}
to define a reduced amplitude 
by factoring off the first two exponential terms:
\be
\hatcA \equiv J^{-1} \exp\left[ - K L \bT_t^2 \right] \cA \,.
\label{defmodamp}
\ee

The removal of $J$ implies that the leading Laurent coefficient 
of $\hatcA$ at $\ell$ loops will be $1/\de^\ell$ rather than $1/\de^{2\ell}$.
The removal of $e^{KL \bT_t^2}$ implies that the leading logarithmic behavior
of $\hatcA$ at $\ell$ loops will be $\log^{\ell-1} |s/t|$.
\para

Note that the reduced amplitude (\ref{defmodamp})
is almost, but not quite the same as, 
the reduced amplitude defined in 
refs~\cite{Caron-Huot:2013fea,Caron-Huot:2017fxr},
which multiplies $\cA$ by a factor  
$J^{-1} \exp[-\alpha_g L \bT_t^2]$
rather than 
$J^{-1} \exp[-K L \bT_t^2]$,
where $\alpha_g$ is the Regge trajectory.
The difference 
$\hat\alpha_g = \alpha_g - K$
vanishes at one-loop order (because we are expanding in $\atil$
rather than $a$) and is IR-finite at two loops \cite{Caron-Huot:2017fxr}.
\para

Using \eqn{finalamp}
we may write the reduced amplitude 
(\ref{defmodamp}) as
\begin{align}
\hatcA
&
=
\exp\left[ i \pi K \bT_{s-u}^2 \right]
\exp\left[ - \half i \pi L  K^2 [ \bT_t^2, \bT^2_{s-u} ] \right]
\label{Ahat}
\\
&
\hskip5mm
\times 
\exp \left[ 
\frac{i \pi}{6} K^3 L^2 [\bT_t^2 ,[\bT_t^2 ,\bT_{s-u}^2 ]] 
- \frac{\pi^2}{3} K^3 L [\bT_{s-u}^2,[\bT_t^2 ,\bT_{s-u}^2]] 
+ \frac{4}{3\de} \atil^3   \Delta^\Three 
\right] 
\cH
+\cO(\atil^4) \,.
\nn
\end{align}

We then expand this in powers of $\atil$, using \eqn{K}, to obtain
\begin{align}
\hatcA
&= 
\cA^\Zero 
+ \left[
 \atil \left( {i \pi \over \de}   \bT_{s-u}^2  \right) \cA^\Zero + \cH^\One 
\right]
\nn\\
&
+ 
\Bigg[
\atil^2 
\bigg(
- {i \pi \zeta_2 \over 2 \de} N   \bT_{s-u}^2 
-{\pi^2 \over 2\de^2}  (\bT_{s-u}^2 )^2 
- {i \pi \over 2 \de^2}    L [ \bT_t^2, \bT^2_{s-u} ] 
\bigg) 
\cA^\Zero
+ \atil \left( {i \pi \over \de}   \bT_{s-u}^2 \right) \cH^\One 
+ \cH^\Two
\Bigg]
\nn\\
&+ 
\Bigg[
\atil^3 
\bigg( \frac{11 i \pi  \zeta_4}{6 \de}N^2\bT_{s-u}^2 
+\frac{\pi ^2   \zeta_2}{2 \de^2}N (\bT_{s-u}^2 )^2 
-\frac{i \pi ^3 }{6 \de^3}(\bT_{s-u}^2 )^3 
+ {i \pi \zeta_2 \over 2 \de^2}  L N  [ \bT_t^2, \bT^2_{s-u} ] 
\nn\\
& 
\hskip5mm
+ \frac{i \pi}{6\de^3}  L^2 [\bT_t^2 ,[\bT_t^2 ,\bT_{s-u}^2 ]] 
- \frac{\pi^2}{3\de^3}  L [\bT_{s-u}^2,[\bT_t^2 ,\bT_{s-u}^2]] 
+  \frac{\pi^2}{2\de^3} L \bT^2_{s-u} [ \bT_t^2, \bT^2_{s-u} ] 
+ \frac{4}{3\de} \Delta^\Three 
\bigg) \cA^\Zero
\nn\\
&
\hskip5mm
+ 
\atil^2 \bigg(
- {i \pi \zeta_2 \over 2 \de} N   \bT_{s-u}^2 
-{\pi^2 \over 2\de^2}  (\bT_{s-u}^2 )^2 
- {i \pi \over 2 \de^2}    L [ \bT_t^2, \bT^2_{s-u} ] 
\bigg) 
\cH^\One
+ \atil \left( 
 {i \pi \over \de}   \bT_{s-u}^2 
\right) \cH^\Two + \cH^\Three
\Bigg]
\nn\\
&
\hskip5mm
+\cO(\atil^4)
\label{IRexpansion} 
\end{align}

where we denote the loop expansion of the hard function as 
\be
\cH
= 
\cA^\Zero + \cH^\One + \cH^\Two + \cH^\Three + \cdots 
\ee

We now consider the amplitude at each order in $\atil$. 

\subsection{One loop}

The $\cO(\atil)$ term of \eqn{IRexpansion} is 
\be
\hatcA^\One = 
\atil \left( {i \pi \over \de}   \bT_{s-u}^2 \right) \cA^\Zero + \cH^\One  \,.
\ee

Recalling that $\cA^\Zero = A_1^\Zero \CX_\one$
and $\CX_\two = \bT_{s-u}^2 \CX_\one$, 
this can be expressed in the Regge color factor basis as
\begin{align}
\hatcA^\One
&= 
A_1^\Zero \atil \left[ 
\hX_\one^\One N \CX_\one
  + 
\left( {i \pi \over  \de}  + \hX_\two^\One \right)
\CX_\two \right]
\label{oneloopIR}
\end{align}

where  we define
\begin{align}
\cH^\One
&= 
A_1^\Zero \atil \left[ \hX_\one^\One   N \CX_\one
                     + \hX_\two^\One   \CX_\two \right] \,.
\end{align}

Using the exact (all orders in $\de$) one-loop amplitude from
\eqns{oneloopreggebasis}{oneloopreggecoefficients},
one obtains the exact one-loop reduced amplitude 
\begin{align}
\hatcA^\One 
&=
A_1^\Zero
\atil 
\left[   
f_1 (\de) N \CX_\one
+
{i \pi \over \de} \CX_\two
\right]  \,.
\label{oneloopexact}
\end{align}

Comparing \eqns{oneloopIR}{oneloopexact}, 
we obtain the one-loop IR-finite contributions 
to all orders in $\de$ :
\begin{align}
\hX_\one^\One &= f_1(\de) 
= \half \pi^2 +  \zeta_3 \de 
+ \fr{1}{30} \pi^4  \de^2 + \zeta_5 \de^3
+ \fr{1}{315} \pi^6 \de^4 + \cdots  \,,
\nn\\[2mm]
\hX_\two^\One &= 0.
\label{Hone}
\end{align}

\subsection{Two loops}

Next, we consider the $\cO(\atil^2)$ term in \eqn{IRexpansion}
\begin{align}
\hatcA^\Two  
&= 
\atil^2 \bigg[
 i \pi  \left( 
 {- \fr{1}{12} \pi^2 + \hX_\one^\One  \over  \de} 
\right)
N \bT_{s-u}^2
+ i \pi \left( - {L \over 2 \de^2}  \right)  [ \bT_t^2, \bT^2_{s-u} ] 
+ \left( -{\pi^2 \over 2\de^2} 
\right)  (\bT_{s-u}^2 )^2 
\bigg] 
\cA^\Zero
+
\cH^\Two \,.
\end{align}

All of the terms in this equation can be expressed 
in terms of the Regge color factor basis 
\begin{align}
\hat \cA^\Two 
&= 
A_1^\Zero \atil^2 \left[ 
\hatBX_\one^\Two N^2    \CX_\one
+
\hatBX_\two^\Two  N \CX_\two
+ 
\hatBX_\three^\Two    \CX_\three
+ 
\hatBX_\four^\Two \CX_\four
\right] \,,
\nn\\
\cH^\Two 
&= 
A_1^\Zero \atil^2 \left[ 
\hX_\one^\Two N^2    \CX_\one
+
\hX_\two^\Two  N \CX_\two
+ 
\hX_\three^\Two    \CX_\three
+ 
\hX_\four^\Two \CX_\four
\right]
\end{align}

where, using \eqn{Hone}, we have 
\begin{align}
\hatBX_\one^\Two
&= 
 \hX_\one^\Two \,,
\nn\\
\hatBX_\two^\Two
&= 
i \pi \left(
 {\fr{5}{12} \pi^2 \over  \de}      
+  \zeta_3  + \fr{1}{30} \pi^4  \de + \zeta_5 \de^2 + \cdots
\right)
+ \hX_\two^\Two \,,
\nn\\
\hatBX_\three^\Two
&=
i \pi \left(  - {L \over 2 \de^2}  \right)   
+ \hX_\three^\Two \,,
\nn\\
\hatBX_\four^\Two
& =
-{\pi^2 \over 2\de^2} 
+ \hX_\four^\Two \,.
\label{hatBTwo}
\end{align}

To determine the two-loop IR-finite contributions $\hX_\ik^\Two$,
we use the ancillary files of Henn and Mistlberger \cite{Henn:2016jdu}
to extract\footnote{In ref.~\cite{Henn:2016jdu}, $s$ is taken to be negative.
To convert their results to our conventions, 
we use the map $s_{HM}=u$, $t_{HM}=t$, $u_{HM}=s$.}
the Regge limit of the two-loop color-ordered amplitudes $A_\lam^\Two$
through $\cO(\de^2)$,
then use \eqns{matrices}{defmodamp}
to derive the corresponding reduced amplitudes $\hatA_\lam^\Two$,
and finally use \eqn{twoloopreggetrace}
to determine $\hatBX_\ik^\Two$ to $\cO(\de^2)$.
In doing so, we recover precisely the IR-divergent terms of \eqn{hatBTwo}
together with the following IR-finite coefficients through $\cO(\de^2)$:
\begin{align}
\hX_\one^\Two 
&= 
- \half \zeta_3 L 
+\fr{7}{360} \pi^4
-  \left( \fr{1}{45}\pi^4 L + \fr{39}{2} \zeta_5 + \fr{5}{12} \pi^2 \zeta_3 \right)\de
+  \left( \fr{41}{2}\zeta_5 L 
+ \fr{3}{2} \pi^2 \zeta_3 L 
-\fr{47}{2} \zeta_3^2 
- \fr{19}{168}\pi^6
\right)\de^2
+ \cO(\de^3)  \,,
\nn\\[2mm]
\hX_\two^\Two 
&=
 i \pi  \left[ 
- \half \zeta_3 
- \fr{1}{45}\pi^4 \de
+ \left( \fr{41}{2} \zeta_5 + \fr{3}{2}\pi^2 \zeta_3 \right)\de^2
\right] 
+ \cO(\de^3)  \,,
\nn\\[2mm]
\hX_\three^\Two
&= 
i \pi \left[ -3 \zeta_3 
- \left(  9 \zeta_3 L + \fr{13}{36}\pi^4 \right) \de
- \left(  \fr{3}{20}\pi^4 L +152 \zeta_5 + \fr{28}{3}\pi^2 \zeta_3 \right)\de^2
  \right]
 + \cO(\de^3)\,,
\nn\\[2mm]
\hX_\four^\Two 
&=
  3 \pi^2 \zeta_3 \de 
+ \fr{1}{20}\pi^6 \de^2 
+ \cO(\de^3) \,.
\label{Htwo}
\end{align}

The coefficients of all the $L$-dependent terms 
of $\hX_\one^\Two$ and $\hX_\three^\Two$ are 
consistent\footnote{Because we are expanding 
in $\atil$ rather than $a$,
the values of the Regge trajectory coefficients 
$\hat\alpha_g^{(\ell)}$ differ from eqs.~(D.1) and (D.2)
of ref.~\cite{Caron-Huot:2017fxr}.  
In particular, $\hat\alpha_g^\One$ vanishes in our case,
as noted in that reference.
\label{reggetraj}
}
with the NLL prediction (4.32) of ref.~\cite{Caron-Huot:2017fxr}.
Also, $\hX_\four^\Two$
is consistent with the NNLL prediction (4.33) 
of ref.~\cite{Caron-Huot:2017fxr},
which in fact allows it to be computed to all orders in $\de$, viz.
\be
\hX_\four^\Two 
= {\pi^2 \over 2 \de^2} 
\left[ 1-  
\frac{\Gamma^2 (1-2 \de) \Gamma (1+2 \de)}
{\Gamma (1 -\de) \Gamma^2 (1+ \de) \Gamma (1- 3 \de)}
\right].
\label{Htwoexact}
\ee
\para

\subsection{Three loops}

Finally we consider the $\cO(\atil^3)$ term in \eqn{IRexpansion}.
Most of the terms can be immediately written 
in terms of the three-loop Regge color basis, with the exception of
$\bT_{s-u}^2 [\bT_t^2, \bT_{s-u}^2]$
and also $\bW$ which appears in $\Delta^\Three$.
Using \eqn{matrices} and \eqn{threeloopcomponents},
we may determine that
\be
\bT_{s-u}^2 [\bT_t^2, \bT_{s-u}^2]   \CX_\one
=
 - \fr{1}{12} N^3 \CX_\one  
+ \fr{1}{3} N (\bT_{s-u}^2 )^2 \CX_\one
+ [\bT_{s-u}^2 [\bT_t^2, \bT_{s-u}^2]  ] \CX_\one \,.
\ee

Using the matrix representation of $\bW$ 
given in eq.~(C.8) of ref.~\cite{Caron-Huot:2017fxr}, 
we find the components of 
$\bW \CX_\one $
in the extended three-loop trace basis
\be
\bW \CX_\one  =  (0,0,0,-1,0, 1, 13,0,-13,0,0,0) \,.
\ee

Then using \eqn{threeloopcomponents}
we can express it 
in terms of the three-loop Regge color factor basis
\be
\bW \CX_\one
= 
- \fr{1}{4} N^3  \CX_\one
+ N (\bT_{s-u}^2 )^2 \CX_\one
- 2 [\bT_{s-u}^2 [\bT_t^2, \bT_{s-u}^2]  ] \CX_\one \,.
\label{Wdecomp}
\ee

We thus obtain  for the three-loop amplitude
\begin{align}
\hatcA^\Three 
&= 
\atil^3 \bigg[ 
\left(
- {\pi^2  \over 24\de^3} L 
- {i \pi \over 12\de}   \hX_\three^\Two 
+ \boxed{ \frac{1}{12 \de} \left( {\zeta_5+2\zeta_2\zeta_3} \right) }
\right) N^3 
\\
&
\hskip5mm
+
\left(
\frac{11 i \pi  \zeta_4}{6 \de}
- {i \pi \zeta_2 \over 2 \de} \hX_\one^\One 
+ {i \pi \over \de} \hX_\one^\Two 
\right)
N^2\bT_{s-u}^2 
+ \left(
  {i \pi \zeta_2 \over 2 \de^2}  L  
- {i \pi \over 2 \de^2}    L \hX_\one^\One 
\right)
N [ \bT_t^2, \bT^2_{s-u} ] 
\nn\\
&
\hskip5mm
+
\left(
\frac{\pi ^2   \zeta_2}{2 \de^2}
-{\pi^2 \over 2\de^2} \hX_\one^\One 
+ {i \pi \over \de}   \hX_\two^\Two  
+{\pi^2  \over 6\de^3} L 
+ {i \pi \over 3 \de}   \hX_\three^\Two 
- 
\boxed{
\frac{1}{3 \de} \left( {\zeta_5+2\zeta_2\zeta_3} \right)
}
\right)
N (\bT_{s-u}^2 )^2 
\nn\\
& 
\hskip5mm
+ 
\left( 
 \frac{i \pi}{6\de^3}  L^2 
+ 
\boxed{
 \frac{i \pi}{3 \de} \left( \zeta_3 L + 11 \zeta_4 \right)
}
\right)
[\bT_t^2 ,[\bT_t^2 ,\bT_{s-u}^2 ]] 
\nn\\
&
\hskip5mm
+
\left( 
  \frac{\pi^2}{6\de^3}  L 
+ {i \pi \over \de}   \hX_\three^\Two 
+ \boxed{
 \frac{1}{\de}  \zeta_5 
}
\right)
[\bT_{s-u}^2,[\bT_t^2 ,\bT_{s-u}^2]] 
+
\left( 
-\frac{i \pi ^3 }{6 \de^3}
+ {i \pi \over \de}   \hX_\four^\Two 
\right)
(\bT_{s-u}^2 )^3 
\Bigg] \cA^\Zero
+ \cH^\Three
\nn
\end{align}

where we have boxed all the terms that come from
the three-loop dipole correction term $\Delta^\Three$.
Expressing the amplitude in the three-loop Regge color factor basis
\begin{align}
\hatcA^\Three 
&=
A_1^\Zero \atil^3 
\Big[ 
\hatBX_\one^\Three N^3  \CX_\one
+
\hatBX_\two^\Three  N^2 \CX_\two
+ 
\hatBX_\three^\Three N \CX_\three
+ 
\hatBX_\four^\Three N \CX_\four
+
\hatBX_\five^\Three \CX_\five
+ 
\hatBX_\six^\Three \CX_\six
+ 
\hatBX_\seven^\Three \CX_\seven
\Big] \,,
\nn\\
\cH^\Three 
&= 
A_1^\Zero \atil^3 
\Big[ 
\hX_\one^\Three N^3  \CX_\one
+
\hX_\two^\Three  N^2 \CX_\two
+ 
\hX_\three^\Three N \CX_\three
+ 
\hX_\four^\Three N \CX_\four
+
\hX_\five^\Three \CX_\five
+ 
\hX_\six^\Three \CX_\six
+ 
\hX_\seven^\Three \CX_\seven
\Big]
\end{align}

and using \eqns{Hone}{Htwo} we obtain 
\begin{align}
\hatBX_\one^\Three 
&= 
\left(
-  {\pi^2  \over 24\de^3} L 
+ \frac{ - \fr{ 2}{9}  \pi^2 \zeta_3 + \frac{1}{12 } \zeta_5 } {\de}
- \fr{ 3}{4}  \pi^2 \zeta_3  L
- \fr{13}{432}\pi^6
\right) 
+ \hX_\one^\Three \,,
\nn\\
\hatBX_\two^\Three 
&= 
i \pi \left(
\frac{
- \fr{1}{540} \pi^4 
- \fr{1}{2} \zeta_3  L  }{\de}
- \fr{1}{45} \pi^4 L - \fr{39}{2} \zeta_5 - \half \pi^2 \zeta_3
\right)
+ \hX_\two^\Three \,,
\nn\\
\hatBX_\three^\Three 
&=
i \pi 
\left(
- { \pi^2  \over 6 \de^2}  L  
- { \zeta_3  \over 2 \de}  L  
- \fr{1}{60} \pi^4 L
\right)
+ \hX_\three^\Three \,,
\nn\\
\hatBX_\four^\Three 
& =
\left(
 {\pi^2  \over 6\de^3} L 
-{ \pi^4  \over 6\de^2} 
+ 
\frac{
 \fr{8}{9} \pi^2 \zeta_3 - \frac{1}{3} \zeta_5 }
{\de}
+ 3 \pi^2 \zeta_3 L + \fr{17}{135} \pi^6
\right)
+ \hX_\four^\Three \,,
\nn\\
\hatBX_\five^\Three 
& =
i \pi \left( 
 \frac{1}{6\de^3}  L^2 
+ \frac{  \fr{1}{3}  \zeta_3 L + \fr{11}{270} \pi^4  }
{\de} 
\right)
+ \hX_\five^\Three \,,
\nn\\
\hatBX_\six^\Three 
&=
\left( 
  \frac{\pi^2}{6\de^3}  L 
+ \frac{ 3\pi^2 \zeta_3 + \zeta_5 }{\de}
+ 9 \pi^2 \zeta_3 L + \fr{13}{36} \pi^6
\right)
+ \hX_\six^\Three \,,
\nn\\
\hatBX_\seven^\Three 
&=
i \pi \left( 
-\frac{ \pi^2 }{6 \de^3}
+ 3 \pi^2 \zeta_3 
\right)
+ \hX_\seven^\Three \,.
\label{hatBThree}
\end{align}

Once again, to determine the IR-finite contributions $\hX_\ik^\Three$,
we use the ancillary files of Henn and Mistlberger \cite{Henn:2016jdu}
to extract the Regge limit of the three-loop color-ordered 
amplitudes $A_\lam^\Three$
through $\cO(\de^0)$,
then use \eqn{defmodamp}
to derive the corresponding reduced amplitudes $\hatA_\lam^\Three$,
and finally use \eqn{threeloopreggetrace}
to determine $\hatBX_\ik^\Three$ to $\cO(\de^0)$.
In doing so, we recover precisely the IR-divergent terms of \eqn{hatBThree}
together with the IR-finite coefficients at $\cO(\de^0)$:
\begin{align}
\hX_\one^\Three
&= 
2 \zeta_5 L
-\frac{11}{36} \pi ^2  \zeta_3 L
-\frac{65 \zeta_3^2}{36}
-\frac{1397 \pi ^6}{102060} + \cO(\de) \,,
\nn\\
\hX_\two^\Three
&= 
i \pi \left(
2 \zeta_5 
-\frac{5 \pi ^2 \zeta_3}{36}
\right) + \cO(\de) \,,
\nn\\
\hX_\three^\Three
&= 
i \pi \left(
\frac{\pi ^4 }{60} L
+2 \zeta_5
-\frac{13 \pi ^2 \zeta_3}{6}
\right) + \cO(\de) \,,
\nn\\
\hX_\four^\Three
&= 
\frac{4}{3} \pi ^2  \zeta_3 L
-\frac{\zeta_3^2}{3}
+\frac{565 \pi ^6}{6804} + \cO(\de) \,,
\nn\\
\hX_\five^\Three
&= 
i \pi \left(
-\frac{11  \zeta_3}{3} L^2
-\frac{29 \pi ^4 }{90} L
-\frac{65 \zeta _5}{6}
+\frac{5 \pi ^2 \zeta_3}{12}
\right) + \cO(\de) \,,
\nn\\
\hX_\six^\Three
&= 
\frac{10}{3} \pi ^2  \zeta_3 L
+\zeta_3^2
+\frac{32 \pi ^6}{567} + \cO(\de) \,,
\nn\\
\hX_\seven^\Three
&= 
i \pi \left(
\frac{2 \pi ^2 \zeta_3}{3}
\right) + \cO(\de) \,.
\label{Hthree}
\end{align}

The coefficient of $L^2$ of $\hX_\five^\Three$ matches the NLL prediction
(4.42) of ref.~\cite{Caron-Huot:2017fxr},
and the coefficients of $L$ of 
$\hX_\one^\Three$,
$\hX_\four^\Three$, and 
$\hX_\six^\Three$ 
match\footnote{See footnote \ref{reggetraj}. Also, 
$D_g^\One = \fr{1}{4} N r(\de) f_1 (\de)$
since we are expanding in $\atil$ rather than $a$.}
the NNLL prediction (4.44) of ref.~\cite{Caron-Huot:2017fxr}.
\para

\subsection{All loop results}

We wish to characterize the logarithmic dependence of
each coefficient $B_\ik^\Ell$ 
in the expansion 

\be
\cA^\Ell
= A_1^\Zero \atil^\ell
\sum_{i=0}^\ell \sum_{k}
\BX_\ik^\Ell N^{\ell-i}   \CX_\ik \,.
\label{reggeexpansionrepeat}
\ee

In the IR-divergent prefactors in \eqn{finalamp}, 
we observe that each power of $L$ requires the 
presence of a factor of $\bT^2_t$.  
The Regge color factor $\CX_\ik$ contains $i-k$
factors of $\bT^2_t$,
and $N^{\ell-i}$ can correspond to $\ell-i$ additional
factors of $\bT^2_t$
(since $\bT^2_t$ acting on $\CX_\one$ produces a factor of $N$).
Hence we anticipate that 
$\BX_\ik^\Ell$ can contain up to $\ell-k$ powers of $L$,
that is 
\be
\BX_\ik^\Ell \sim L^{\ell-k} + \hbox{ lower logarithmic terms}.
\label{expectation}
\ee

Thus $\BX_\one^\Ell$ alone contributes at leading log (LL) order,
$\BX_{i1}^\Ell$ starts at NLL, $\BX_{i2}^\Ell$ at NNLL, etc.
We cannot make this argument completely rigorous,
because {\it a priori} we have no knowledge of the 
$L$ dependence of the hard function $\cH$, 
but the explicit results through three loops
presented earlier are in accord with our expectations.
(The $L$ dependence of the reduced amplitude coefficients
$\hatBX_\ik^\Ell$ is sometimes milder than \eqn{expectation}
because we have stripped off some of the logarithms
in \eqn{defmodamp}.)
\para

By \eqn{expectation}, 
the coefficient $\BXL$ should have no logarithmic dependence
whatsoever.  
In fact, the only term in \eqn{finalamp} that contributes to
$\BXL$ is

\be
\cA \sim \exp\left[ { i \pi \atil  \over \de}  \bT_{s-u}^2 \right] 
\cH
\ee
hence
\be
\cA^\Ell \sim
{1 \over \ell!} \left( {i \pi \atil \over \de} \bT^2_{s-u} \right)^\ell \cA^\Zero
+ \hbox{ contributions from } \cH^\Two, \cH^\Three, \cdots
\ee

and thus
\be 
\BXL 
= 
{1 \over \ell!} 
\left( {i \pi \over \de} \right)^\ell 
+ \cO(\de^{3-\ell})
\label{IRBXL}
\ee

where the $\cO(\de^{3-\ell})$ corrections come from 
$\cH^\Two$, $\cH^\Three, \cdots$.
This is consistent with the explicit one-, two-, and 
three-loop results\footnote{Note that 
$\BXL = \hatBX_{\ell\ell}^\Ell$
because the prefactors in \eqn{defmodamp} 
cannot contribute to this coefficient.}
(\ref{oneloopreggecoefficients}), 
(\ref{hatBTwo}), 
(\ref{Htwoexact}), 
(\ref{hatBThree}), and
(\ref{Hthree})
\begin{align}
B^\One_\two &= {i \pi \over \de} \,,
\nn\\[1mm]
B^\Two_\four &= 
\left( - {\pi^2 \over 2 \de^2}  \right)
\frac{\Gamma^2 (1-2 \de) \Gamma (1+2 \de)}
{\Gamma (1 -\de) \Gamma^2 (1+ \de) \Gamma (1- 3 \de)} \,,
\nn\\[1mm]
B^\Three_\seven &= i \pi \left(  - { \pi^2 \over 6 \de^3}
+ \frac{11\pi^2 \zeta_3 }{3} \right) + \cO(\de) \,.
\label{BXL}
\end{align}
In sec.~\ref{sec:grav}, 
we will observe that 
the coefficients $\BXL$
are closely related to the $\cN=8$ supergravity four-point amplitude 
in the Regge limit.

\section{SYM/supergravity relation in the Regge limit}
\setcounter{equation}{0}
\label{sec:grav}

In this section, 
we review the Regge limit of the $\cN=8$ four-point amplitude.
This will allow us to make an all-loop-orders conjecture
between the 
$\cN=4$ SYM and $\cN=8$ supergravity four-point amplitudes
in the Regge limit.
\para

The tree-level four-graviton amplitude is\footnote{In
this paper, we 
take $G$ to be the four-dimensional Newton's constant,
with $G_D = G \mu^{2 \de}$ its counterpart in $D=4-2 \de$
dimensions.
}\cite{Bern:1998ug}
\be
\cM^\Zero = 
8 \pi G \mu^{2\de} {16 K \tilde{K}  \over s t u } \,.
\ee
where $K$ is defined as in \eqn{treelevelcolorordered}.
The one-loop 
$\cN=8$ supergravity four-graviton amplitude is \cite{Bern:1998ug}
\begin{align}
\cM^\One 
&= 
- i 
(8 \pi G)^2  \mu^{2\de} 
(16 K \tilde{K} ) 
\left[   \cI^\One (s,t) + \cI^\One (u,s) + \cI^\One (t,u) \right]
\nn\\[2mm]
&=
\cM^\Zero \Big( 
- 8 \pi i  G
\, 
stu
\left[   \cI^\One (s,t) + \cI^\One (u,s) + \cI^\One (t,u) \right]
\Big)
\end{align}

which, using the expression of $\cI^\One$ given in \eqn{oneloopintegral}
becomes 
\begin{align}
\cM^\One
=
\cM^\Zero 
\etatil \,  {1 \over \de^2} 
&\Big[ 
(x-1) \left(  \e^{i \pi } x  \right)^{\de} F\left(\de, 1-{1\over x}\right)
+ (x-1) F\left(\de, 1-x  \right)  
\label{oneloopgrav}
\\
&
- x \left(  \e^{i \pi } x \right)^{\de} F\left(\de, {-x \over 1-x}\right)
- x \left( x \over 1-x \right)^{\de} F\left(\de, x  \right) 
\nn\\
&
+ F\left(\de, {1\over 1-x}\right)
+ \left( x \over 1-x \right)^{\de} F\left(\de, {1 \over x} \right) 
\Big]
\nn
\end{align}

where we define the dimensionless parameter\footnote{Related
by $\etatil = \alpha_G s/(-t)^\de$ to $\alpha_G$ defined in
refs.~\cite{DiVecchia:2019myk,DiVecchia:2019kta}.}
\be
\etatil
\equiv  
{G s \over  \pi} 
{ \Gamma^2(1-\eps) \Gamma(1+\eps) \over \Gamma(1-2\eps)} 
\left( 4 \pi \mu^2  \over -t \right)^{\de}  \,.
\ee

Writing the all-orders amplitude as 
\be
\cM = \cM^\Zero \left[ 1 + \suml \etatil^\ell M^\Ell \right]
\ee
we see that 
\eqn{oneloopgrav} simplifies dramatically in the Regge limit $x \to 0$ to 
\be
M^\One = ~\,-\, { i \pi  \over \de } + \cO(x) \,.
\ee

Using eikonal exponentiation in impact parameter space,
the authors of ref.~\cite{DiVecchia:2019myk,DiVecchia:2019kta}
showed that the Regge limit of the $\ell$-loop four-graviton amplitude 
is given to all orders in $\de$ by
\be
M^\Ell  = {1 \over \ell!} \left( 
-\, { i \pi  \over \de } \right)^\ell G^\Ell (\de) 
+ \cO(x)
\label{Mell}
\ee

where
\begin{align}
G^\Ell (\de) 
&= 
\frac{\Gamma^\ell (1-2 \de) \Gamma (1+\ell \de)}
{\Gamma^{\ell-1} (1 -\de) \Gamma^\ell (1+ \de) \Gamma (1- (\ell+1) \de)}
\nn\\[2mm]
&
 = 1 -\fr{1}{3} \ell \left(2 \ell^2+3 \ell-5\right) \zeta_3 \de^3
+ \cO (\de^4).
\end{align}

In particular, \eqn{Mell} gives  (in the Regge limit)
\begin{align}
M^\One 
&= 
 ~\,-\, { i \pi  \over \de } + \cO(x) \,, \nn\\
M^\Two 
&= 
\left( - {\pi^2 \over 2 \de^2}  \right)
\frac{\Gamma^2 (1-2 \de) \Gamma (1+2 \de)}
{\Gamma (1 -\de) \Gamma^2 (1+ \de) \Gamma (1- 3 \de)} + \cO(x) \,,
\nn\\
M^\Three
&=  i \pi \left( {\pi^2 \over  6\de^3} 
- \frac{11 \pi^2 \zeta_3 }{3}
- \frac{11 \pi^6 }{180} \de \right) + \cO(\de^2) 
+ \cO(x) \,. 
\label{Mtwoandthree}
\end{align}

We observe that (up to signs) these are precisely 
the one-, two-, and three-loop values of 
the coefficients $\BXL$ of the $\cN=4$ SYM four-gluon amplitude
given in \eqn{BXL}:
\be
\BX_\two^\One = -  M^\One + \cO(x) \,, \qquad
\BX_\four^\Two =   M^\Two + \cO(x) \,, \qquad
\BX_\seven^\Three = -  M^\Three + \cO(\de) + \cO(x) 
\ee

where the one- and two-loop relations hold to all orders in $\de$,
and the three-loop relation holds 
to the accuracy of the $\cN=4$ SYM calculation.
This motivates the all-loop-orders conjecture:
\be
\boxed{
\BX_{\ell \ell}^\Ell
=
(-1)^\ell M^\Ell  + \cO(x)
}
\label{allordersrelation}
\ee

relating the Regge limits 
of the $\cN=8$ supergravity amplitude
and the $\cN=4$ SYM amplitude.  
Comparing the leading infrared-divergent contribution (\ref{IRBXL})
with \eqn{Mell} 
confirms that \eqn{allordersrelation} is valid
for any $\ell$ to at least the first three orders in 
the Laurent expansion in $\de$.
Based on all this, 
we conjecture that \eqn{allordersrelation}
holds to all orders in $\de$ for any $\ell$. 
\para

We can express the SYM/supergravity relation (\ref{allordersrelation}) 
directly in terms of color-ordered amplitudes
using eqs.~(\ref{evenBXL}-\ref{oddBXL})
\begin{align}
M^\One
& 
= {A_1^{(1,0)} +A_3^{(1,0)}  \over A_1^\Zero \atil } + \cO(x) \,,
&& \ell = 1 \,,
\nn\\
M^\Ell
&= 
{ \fr{1}{2} \left( A_1^{(\ell,\ell)} -A_3^{(\ell,\ell)} \right) 
\over 
 3^{\ell/2} 
\cdot A_1^\Zero \atil^\ell } + \cO(x) \,,
&& \hbox{even } \ell\,,
\nn\\
M^\Ell
&= 
\frac{
-           \left( A_1^{(\ell,\ell-3)} +A_3^{(\ell,\ell-3)} \right)
+ \fr{1}{4} \left( A_4^{(\ell,\ell-2)} +A_6^{(\ell,\ell-2)}
+ A_1^{(\ell,\ell-1)} +A_3^{(\ell,\ell-1)} \right) 
	}
{  3^{(\ell-1)/2} \cdot A_1^\Zero \atil^\ell  }  + \cO(x) \,,
&&
\hbox{odd } \ell > 1 \,.
\label{repeatrelation}
\end{align}

The validity of \eqn{repeatrelation}
for $\ell=1$ and $\ell=2$ is not really surprising since 
they are the Regge limits of more general exact relations.
The one-loop relation is the Regge limit 
of the exact one-loop relation
\cite{Green:1982sw,Naculich:2008ys}
\be
{\cM^\One \over \cM^\Zero  \etatil}
= (1-x) 
{\left( A_1^{(1,0)} +A_2^{(1,0)} + A_3^{(1,0)}\right)  
\over A_1^\Zero \atil }
\label{exactone}
\ee
and the $\ell=2$ relation 
is the Regge limit of the
more general exact two-loop relation \cite{Naculich:2008ys}
\be
{\cM^\Two \over \cM^\Zero  \etatil^2}
=  
\left( 1-x \over 6 \right)
{\left[ (1-x) A_1^{(2,2)} + x A_2^{(2,2)} - A_3^{(2,2)} \right] 
\over A_1^\Zero \atil^2 }
\label{exacttwo}
\ee

where we have expressed both relations 
using the notation of the current paper.
Both \eqns{exactone}{exacttwo} are proved by expressing
the amplitudes in terms of planar and nonplanar integrals
\cite{Bern:1997nh,Bern:1998ug,Naculich:2008ys}.
Although previous attempts at 
finding exact SYM/supergravity relations
beyond two loops were not successful, 
there may nevertheless exist exact relations
of which \eqn{repeatrelation} 
is the Regge limit.

\section{Conclusions}
\setcounter{equation}{0}
\label{sec:concl}

We presented in this paper an all-loop-order basis 
of color factors $\CX_\ik$
suitable for writing the Regge limit of the $\cN=4$ SYM four-gluon amplitude.
These color factors have well-defined signature 
under crossing symmetry $u \leftrightarrow s$;
specifically,  $\CX_\ik$ has negative/positive signature for $k$ even/odd.
\para

We found that the coefficients $\BX_\ik^\Ell$ 
of the Regge limit of the $\cN=4$ SYM amplitude in this basis 
are polynomials of order $\ell-k$ in $L = \log |s/t| - \half i \pi$. 
Thus $\BX_\one^\Ell$ alone contributes at leading log (LL) order,
$\BX_{i1}^\Ell$ starts at NLL, $\BX_{i2}^\Ell$ at NNLL, etc.
The coefficients of color factors with negative/positive signature
are real/imaginary respectively (when expressed in terms of $L$),
as shown on general grounds in ref.~\cite{Caron-Huot:2017fxr}.
Using results from ref.~\cite{Henn:2016jdu},
we computed these coefficients explicitly through three-loop order,
verifying consistency with all the expected IR divergences 
\cite{Catani:1998bh,Sterman:2002qn,Bern:2005iz,
Aybat:2006wq,Aybat:2006mz,
Dixon:2008gr,Becher:2009cu,Gardi:2009qi,Dixon:2009gx,
Becher:2009qa,Gardi:2009zv,Dixon:2009ur,Almelid:2015jia,
Bret:2011xm,DelDuca:2011ae,Caron-Huot:2013fea,
DelDuca:2013ara,DelDuca:2014cya,Caron-Huot:2017fxr},
as well as with certain IR-finite NLL and NNLL predictions
\cite{Caron-Huot:2013fea,Caron-Huot:2017fxr}.
\para

Based on our explicit results,
we conjectured an all-loop-orders equivalence (up to sign) 
between the coefficients $\BXL$ and the Regge limit of
the $\ell$-loop $\cN=8$ supergravity four-point amplitude.
This equivalence was proven to be valid to all orders in $\de$
at one and two loops, 
through $\cO(\de^0)$ at three loops, 
and for the first three terms in the Laurent expansion
in $\de$  at $\ell$ loops. 
\para

Naturally, it would be nice to establish the validity  
of the conjectured SYM/supergravity relation 
more generally,
perhaps via an eikonal exponentiation approach 
\cite{DiVecchia:2019myk,DiVecchia:2019kta},
or via known representations of these amplitudes
in terms of planar and nonplanar 
integrals \cite{Bern:1997nh,Bern:1998ug,Bern:2007hh,Bern:2008pv,Bern:2010tq,
Bern:2012uf,Bern:2012uc,Bern:2012uf,Bern:2017ucb}.
\para

It would also be interesting to know if the SYM/supergravity relations 
for $\ell > 2$
are the Regge limits of more general exact SYM/supergravity relations,
as is the case for $\ell=1$ and $\ell=2$.
\para

Finally,
it would be intriguing to discover 
all-orders-in-$\de$ expressions for the
other coefficients $\BX_\ik^\Ell$ of the Regge basis of color factors.

\section*{Acknowledgments}
This material is based upon work supported by the
National Science Foundation under Grant No.~PHY17-20202.

\end{document}